\documentclass[preprint,journal]{vgtc}            

\onlineid{1158}

\ieeedoi{10.1109/TVCG.2023.3327150}

\vgtccategory{Research}

\vgtcpapertype{System}
\usepackage{caption,subcaption,enumitem,amssymb,wasysym}
\PassOptionsToPackage{hyphens}{url}
\title{\toolname{}: A Tool for Guiding Chart and Caption Emphasis}

\author{%
  \authororcid{Dae Hyun Kim}{0000-0002-8657-9986},
  \authororcid{Seulgi Choi}{0000-0002-9334-0471}, 
  \authororcid{Juho Kim}{0000-0001-6348-4127},
  \authororcid{Vidya Setlur}{0000-0003-3722-406X}, and 
  \authororcid{Maneesh Agrawala}{0000-0002-8996-7327}
}

\authorfooter{
  \item
  	Dae Hyun Kim is with KAIST. This work began when Dae Hyun was at Stanford University.
  	E-mail: dhkim16@cs.stanford.edu
  \item
        Seulgi Choi is with KAIST.
        E-mail: igules8925@kaist.ac.kr
  \item 
        Juho Kim is with KAIST.
        E-mail: juhokim@kaist.ac.kr
  \item 
        Vidya Setlur is with Tableau Research.
        E-mail: vsetlur@tableau.com
  \item 
        Maneesh Agrawala is with Stanford University and Roblox.
        E-mail: maneesh@cs.stanford.edu
}

\abstract{
    Recent work has shown that when both the chart and caption emphasize the same aspects of the data, readers tend to remember the doubly-emphasized features as takeaways; when there is a mismatch, readers rely on the chart to form takeaways and can miss information in the caption text. Through a survey of 280 chart-caption pairs in real-world sources (e.g., news media, poll reports, government reports, academic articles, and Tableau Public), we find that captions often 
do not emphasize the same information in practice, which could limit how effectively readers take away the authors' intended messages.
Motivated by the survey findings, we present \toolname{}, an interactive tool that highlights visually prominent chart features as well as the features emphasized by the caption text along with any mismatches in the emphasis. The tool implements a time-series prominent feature detector based on the Ramer-Douglas-Peucker algorithm and a text reference extractor that identifies time references and data descriptions in the caption and matches them with chart data.
This information enables authors to compare features emphasized by these two modalities, quickly see mismatches, and make necessary revisions. A user study confirms that our tool is both useful and easy to use when authoring charts and captions.
}


\graphicspath{{figs/}{figures/}{pictures/}{images/}{./}} 

\usepackage{tabu}                      
\usepackage{booktabs}                  
\usepackage{lipsum}                    
\usepackage{mwe}                       

\usepackage{mathptmx}                  

\newcommand{\toolname}[0]{\textsc{EmphasisChecker}}

\begin{document}


\firstsection{Introduction}

\maketitle

\begin{figure}
  \centering
  \includegraphics[width=0.9\linewidth]{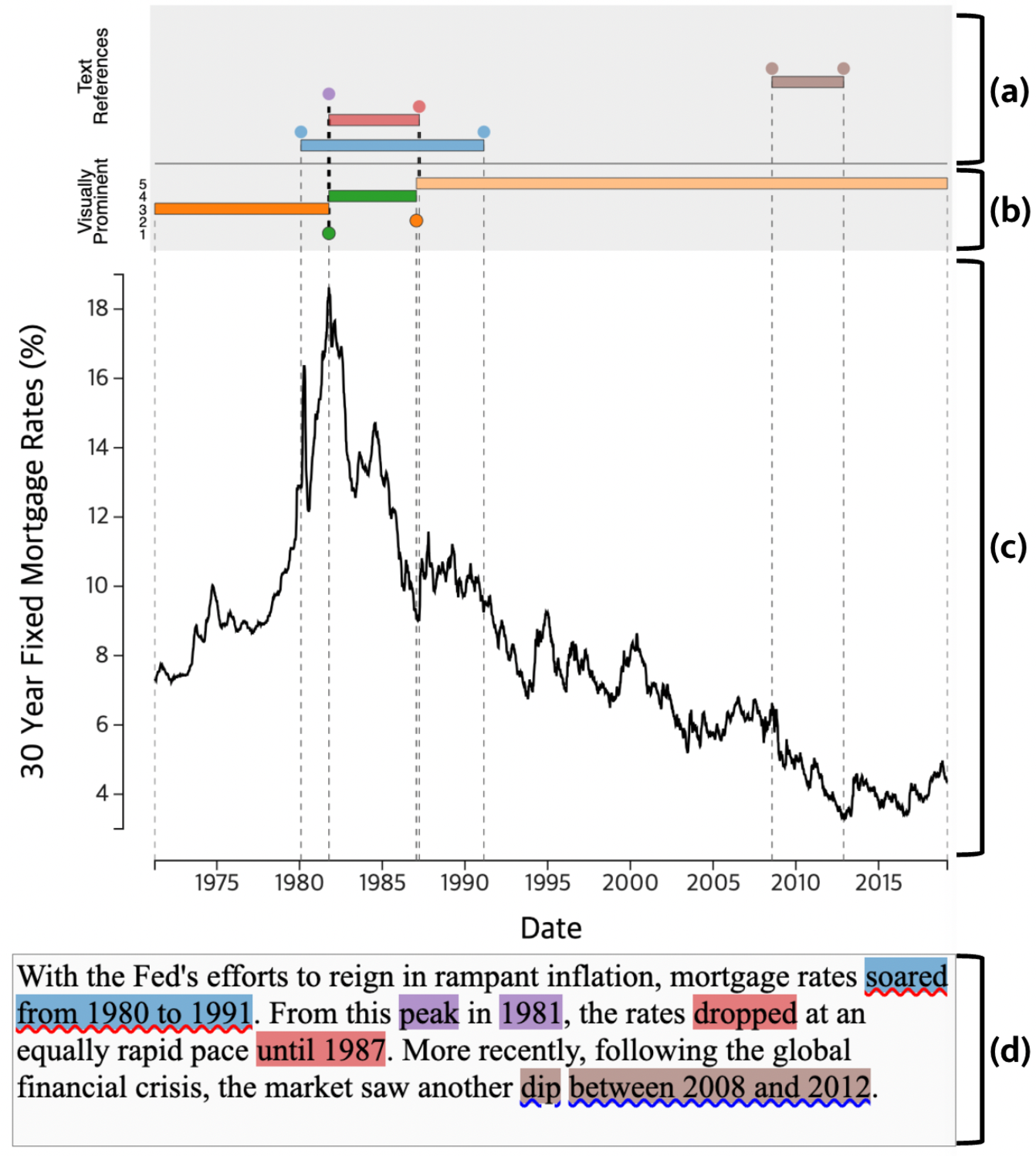}
  \caption{%
    As the author writes a caption about the chart (c) in the textbox (d), \toolname{} shows the chart's visually prominent features in (b) and the text references to the chart features (a). The interface shows (b) visually prominent chart features (unmatched features in orange {\color[RGB]{227,126,36}$\blacksquare$} and matched features in green {\color[RGB]{97,158,58}$\blacksquare$}, marks above the chart). 
   It uses circles to depict point features (e.g., local extrema; the peak around 1981) and bars to depict trend features (e.g., the rising trend up to 1981).
   In addition, it shows (a) references between the chart and the text (i.e., blue {\color[RGB]{138,174,210}$\blacksquare$}, red {\color[RGB]{199,116,118}$\blacksquare$}, purple {\color[RGB]{163,140,196}$\blacksquare$}, and brown {\color[RGB]{171,148,143}$\blacksquare$} marks at the top of the page and on the text). In the input text box (d), the tool adds a red squiggly underline (\includegraphics[height=3px]{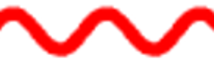})
   on the phrase 
   \textit{`soared from 1980 to 1991'}, a typo of \textit{`soared from 1980 to \textbf{1981}'}
   because the phrase does not match the data in the chart. The tool also adds a blue squiggly underline (\includegraphics[height=3px]{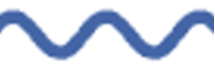}) on the phrase 
   \textit{`dip between 2008 and 2012'}
   because the phrase does not match any of the prominent chart features. 
  }
  \label{fig:teaser}
\end{figure}

Authors often pair charts and caption text together to convey information about data (e.g., in news articles, academic papers, and reports). For example, in Figure~\ref{fig:teaser}, the peak around 1981 is visually prominent. The caption text also emphasizes portions of the data by referring to specific points or ranges of data values. The caption text \textit{`peak in 1981'} emphasizes the visually prominent peak. When the chart and caption text emphasize the same aspects of the data, as in this example, people tend to remember those aspects of the data as takeaways~\cite{cheng2022captions,kim2021towards,xiong2019curse}; when they emphasize different aspects of the data (e.g., the caption text emphasizes the \textit{`dip between 2008 and 2012'}, but the fall is not visually prominent in the chart), people remember the visually prominent data features in the chart, not the data emphasized in the text~\cite{cheng2022captions,kim2021towards}
and question the credibility of the information in the chart and text~\cite{kong2019trust}.

But how often do authors really create charts and caption text that emphasize the same aspects of data? Analyzing time-series charts and their captions in real-world publications (Section~\ref{sec:survey}), we find that professional authors match chart and caption emphasis about 65\% of the time and mismatches in the remaining 35\%.
In a survey of chart-caption pairs on Tableau Public~\cite{tableau2022public}, a community for the general public to share visualizations created using Tableau Software~\cite{tableau2022tableau}, we find that mismatches are even more common among the general public.

To help authors convey their messages effectively, we present \toolname{}, a caption-and-chart checker tool that takes time-series charts
as input. The tool highlights the visually emphasized chart features and the data features emphasized in the caption. The user can then decide how to update the caption or the chart to better align their emphasis.
Our tool follows the model of spell- and grammar-checkers~\cite{google2022check,grammarly2022grammarly,language2022language}, facilitating the process of locating potential mismatches between chart and caption text,
while leaving the final decision of how to resolve the issue to the author. 
In this work, we focus on time-series line charts as a first step towards more general tools as they are among the most common type of charts on the Web~\cite{battle2018beagle}.

Our tool includes a \textit{time-series prominent feature detector} that identifies the visually prominent features of the chart. It also includes a \textit{text reference extractor} that identifies data emphasized in the text. The \toolname{}  interface is designed so that users can then compare the chart-emphasized data features with the text-emphasized data features (Figure~\ref{fig:teaser}).
For the prominent feature detector, we introduce a new $\varepsilon$-persistence technique based on the Ramer-Douglas-Peucker line simplification algorithm for approximating visual prominence. 
The chart-text reference extractor uses heuristics based on example analysis, 
modules within the Stanford CoreNLP toolkit~\cite{chang2012sutime,chen2014fast,finkel2005incorporating,manning2014stanford}, and BERT embeddings~\cite{devlin-etal-2019-bert}
to find time references and data descriptions in the caption text and matches them with chart data.
\toolname~visualizes the results of these components to the user as a part of a chart-caption authoring interface shown in Figure~\ref{fig:teaser}.

We evaluated the two components using the charts from Kim et al.~\cite{kim2021towards}.
We find that our time-series prominent feature detector outperforms a state-of-the-art method for prominent feature detection~\cite{hullman2013contextifier} while performing at the level of crowdsourced prominent features~\cite{kim2021towards}.
For human-written captions on the real-world charts in the corpus, we find that our text reference extractor correctly identifies text references to charts for 63.41\% of the sentences.
Finally, through a user study, we find that \toolname{} is both useful and easy to use when authoring charts and captions.

In summary, the contributions of this work include: 
\begin{itemize}[nolistsep]
    \item a survey of what charts and their captions in the real-world emphasize;
    \item \toolname{}, a caption-and-chart checker tool that guides authors to create charts and captions whose emphasis match chart emphasis; and
    \item algorithms for detecting visually prominent features in time-series line charts and text references to chart features. 
\end{itemize}
\section{Related Work}

Our work is related to two main areas of prior work: 
(1) how people read charts and text and (2) supporting document authoring.
\subsection{How People Read Charts and Text}
Prior research has shown that visualizations and text complement each other when communicating information about the data.
Specifically, Elzer et al.~\cite{elzer2005exploring} and Carberry et al.~\cite{carberry2006information} found that visualizations often contribute information not available in the text alone. On the other hand, researchers have found that text guides readers' attention while viewing visualizations~\cite{gould1976looking,tufte2001visual,xiong2019curse} and improves memorability of the key messages in visualizations~\cite{borkin2015beyond,kim2021towards,cheng2022captions}.

The two representations facilitate different tasks for readers; Ottley et al.~\cite{ottley2019curious} found that readers easily \textit{identify} key information with visualizations and easily \textit{extract} key information from text.
Despite the complementary nature of visualizations and text, prior studies did not find particularly synergistic benefits in comprehending information present in both chart and caption text~\cite{khan2015benefits,micallef2012assessing,ottley2012visually,ottley2015improving}.

Researchers have noticed that one of the major problems limiting the synergistic benefits of using visualizations and text together is \textit{split attention} as readers have to look back and forth between the spatially separated representations~\cite{biderman1979graph,sweller2011split,tufte2001visual}.
Whitacre and Saul~\cite{whitacre2016high} and Ottley et al.~\cite{ottley2019curious} concluded that readers tend not to integrate the information between the visualizations and text. The research community has put forth a range of solutions for the split attention problem. Tufte~\cite{tufte2001visual} introduced \textit{sparklines}, which are word-sized line charts embedded into the text. Goffin et al.~\cite{goffin2015design} and Beck and Weiskopf~\cite{beck2017word} extended the idea by adding interactivity to the sparklines. In addition, multiple systems (e.g., table-text reference display systems~\cite{badam2018elastic,kim2018facilitating}, visualization-text reference display systems~\cite{kong2014extracting,lai2020automatic,latif2018exploring,pinheiro2022charttext})
display the references between visualizations and text to reduce the effort needed to locate relevant information.

The research community has recently begun diving deeper into the relationship between visualizations and text by studying how 
comprehension depends on the content of the visualizations and text.
Kong et al.~\cite{kong2018frames,kong2019trust} studied how slants, framing in chart titles, as well as misalignments between charts and text, affect what people read from charts and their titles. They found that while people identify bias, they still consider charts to be impartial. Whitacre and Saul~\cite{whitacre2016high} studied high school students and found that they had difficulty identifying the inconsistencies between graphs and their captions. Kim et al.~\cite{kim2021towards} found that when charts and captions emphasize the same feature in the data, people tend to take away the message related to the doubly emphasized feature. When the caption describes a feature that is not visually prominent in a chart, readers are more likely to consider the visually prominent chart features as carrying the key messages. Subsequent work found that semantic levels of text~\cite{cheng2022captions,lundgard2021accessible,stokes2022striking}, as well as its placement~\cite{stokes2022striking}, can influence how readers integrate charts with text.
Although our work on the \toolname{}~tool is targeted toward authors, the principles behind how 
the tool helps authors write effective chart-text pairs, rely on these theories about how readers read charts and text together. 

\subsection{Supporting Document Authoring}
Document authoring tools often support authors so that they can easily write high-quality text with a clear exposition.
Spell-checkers~\cite{earnest2011first} have been incorporated into many writing environments, including word processors and e-mail~\cite{google2022check,libreoffice2022checking,google2022fix,microsoft2022checking}.
Grammar checkers are also available in such environments through software extensions or plug-ins, such as Grammarly~\cite{grammarly2022grammarly} and LanguageTool~\cite{language2022language}.
More recently, writing environments have incorporated AI-based autocompletion~\cite{google2022using} that tries to  predict what the writer intends to write.
Similar to these tools, \toolname{} is designed to help people write documents, but instead of purely analyzing text, \toolname{} considers chart and text pairs and analyzes the visuals as well as the words together. 

The research community has developed methods for automatically generating caption text for visualizations. 
Many of these techniques  
are designed to generate basic captions that simply explain how the visualization encodes data (e.g., \cite{cui2019datasite,demiralp2017foresight,tableau2022tableau,wills2010autovis}).
A few techniques go beyond basic captions and generate descriptions and summaries of features (e.g., ranking, extrema, trends) in the visualizations, such as techniques based on Bayesian models~\cite{carberry2006information,elzer2011automated} and neural networks~\cite{chen2019neural,chen2019figure,obeid2020chart,qian2021generating}.
Rather than generating captions, Contextifier~\cite{hullman2013contextifier} adds annotations based on financial news headlines to noteworthy features of stock charts.
Unlike the fully-automated systems, \toolname~follows the interface paradigm of a spell- or grammar-checker by guiding the author while they have complete control over their writing.

More advanced tools provide assistance beyond adding text to visualizations. TimeLineCurator~\cite{Fulda2016TimeLineCuratorIA} is a web-based timeline authoring tool that automatically extracts event data from temporal references in unstructured text documents using natural language processing, along with controls for curating and editing the events. VizByWiki~\cite{lin2018vizbywiki} retrieves visualizations relevant to given news articles from Wikimedia Commons~\cite{wikimedia2022wikimedia} to enrich articles.
PostGraphe~\cite{fasciano1996postgraphe} creates text and graphics based on tabular input data and users' intents.
Kori~\cite{latif2021kori} takes a mixed-initiative approach in helping authors construct interactive references between visualizations and text into their documents with suggestions based on natural language processing techniques.
CrossData~\cite{chen2022crossdata} links text and the underlying data to reduce efforts in writing data documents and data exploration.
Although \toolname{} leverages references between visualizations and text similarly to these prior works, our tool also analyzes the \textit{chart data} while identifying references in order to establish comparisons between chart and text emphasis. Furthermore, the reference extraction is only a component of \toolname{} whose end goal is to provide an interface for identifying mismatches between visualizations and the text.

\section{A Survey of Line Charts in the Wild}
\label{sec:survey}
To understand whether charts and captions emphasize the same information in practice, we conducted a survey of chart-text pairs in various real-world sources.
We specifically looked at charts and captions written by professionals and circulated through various publishing venues, as well as charts and captions published on Tableau Public~\cite{tableau2022public}, a community of professional and non-professional authors for sharing charts and captions created with Tableau Software~\cite{tableau2022tableau}, a tool that suggests basic captions about data fields and chart encoding to its users.

\subsection{Dataset}

For chart-caption pairs written by authors, we sampled a total of 250 chart-article pairs (189 unique articles) from various publishing venues to obtain a representative set.
The venues include news media (New York Times~\cite{nyt2022nyt}, The BBC~\cite{bbc2022bbc}, Vox~\cite{vox2023vox}), poll reports (Pew Research~\cite{pew2022pew}), governmental and intergovernmental organization reports (US Treasury~\cite{us2023treasury}, International Monetary Fund~\cite{imf2023imf}, International Labour Organization~\cite{un2023ilo}, etc.), and academic articles (Nature~\cite{springer2023nature}).
In addition, we sampled 30 chart-article pairs (21 unique charts) from Tableau Public. We programmatically scraped the Web for articles published up to two years from the collection date and filtered for line charts using the chart classification pipeline from Poco and Heer~\cite{poco2017reverse}.
Because Pew Research did not allow scraping without permission, we performed the collection process manually for that site.
We include the links to these articles in the supplemental material.

\subsection{Analysis Method}
We performed analysis on the chart-article pairs in two steps; we identified (1) chart emphasis and then (2) text emphasis.

\vspace{0.05in}
\noindent\textbf{\textit{Step 1: Identify chart emphasis.}}
For each chart image in the dataset, two of the authors of this paper independently labeled the visually prominent features.
Afterward, the authors merged their labels and whenever the labels did not match, the two authors shared their reasoning for their annotations with one another to arrive at a consensus. If no consensus was reached, the authors used the opinion of a third individual who works in visualization research but is not a co-author to determine the prominent features of the chart. At this stage, the authors did not read any of the text of the article to avoid biasing their perception of visually prominent features~\cite{xiong2019curse}.

\vspace{0.05in}
\noindent\textbf{\textit{Step 2: Identify text emphasis.}}
The two authors read through the article text to identify all paragraphs that mention any information about the chart by looking for mentions of the phrases in the chart title, axis names, or chart numbering (e.g., `\textit{Figure X},' `\textit{Chart X}').
They each independently perused the sentences in each of the paragraphs, chart titles, and textual annotations within the charts to identify any references to prominent features and non-prominent features.
The authors again discussed the annotations to arrive at a consensus, using the third individual for arbitration when necessary to resolve conflicts.

\subsection{Results}
\begin{figure}
    \centering
    \includegraphics[width=\linewidth]{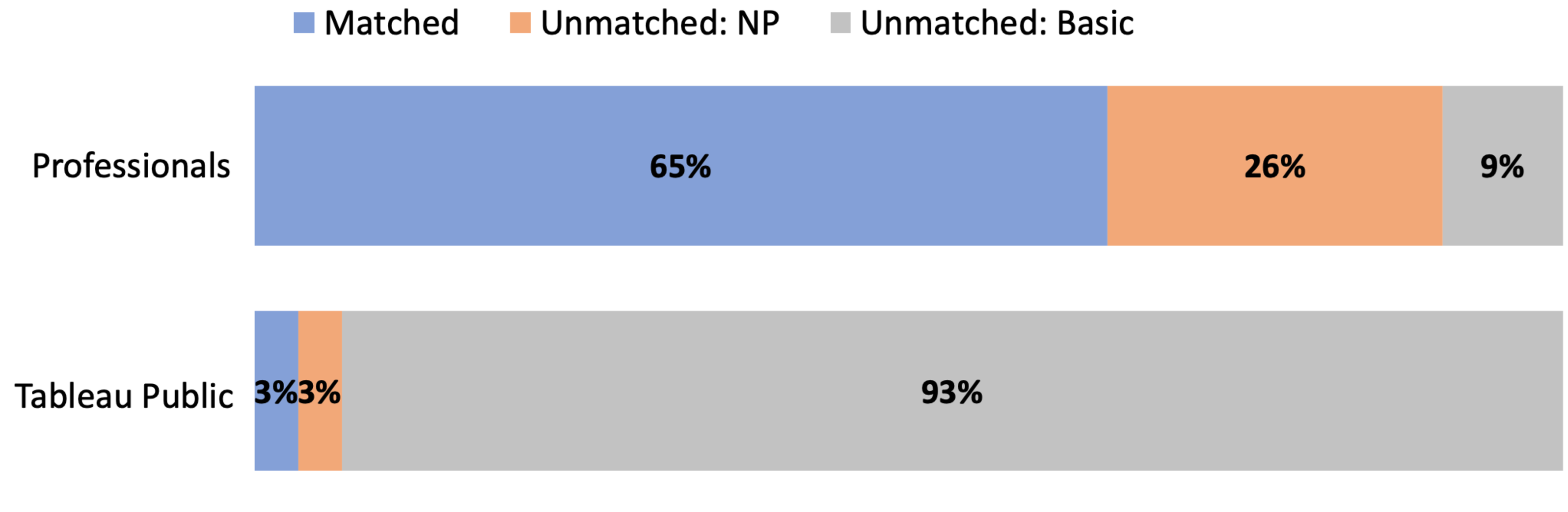}
    \caption{Distribution of chart and caption emphasis in chart-article pairs.
    Professionals often match chart and text emphases (blue {\color[RGB]{140, 158, 212}$\blacksquare$} segment in the top bar) but occasionally do not (orange {\color[RGB]{224, 164, 122}$\blacksquare$} (Unmatched: NP (non-prominent); captions only describing non-prominent chart features) + gray {\color[RGB]{192, 192, 192}$\blacksquare$} (Unmatched: basic; basic captions that do not point to specific chart features) segments in the top bar).
    On the other hand, captions on Tableau Public are predominantly basic captions (gray {\color[RGB]{192, 192, 192}$\blacksquare$} segment in the bottom bar).
    The values have been rounded to the nearest whole numbers and may not sum to 100\%.}
    \label{fig:survey-results}
    \vspace{-0.15in}
\end{figure}

Figure~\ref{fig:survey-results} shows the results of our analysis.
We observed a visible dividing line between chart captions written by professional authors circulated through publishing organizations and chart-caption pairs on Tableau Public.

\vspace{0.05in}
\noindent\textbf{\textit{Professionals often match emphasis but occasionally do not.}}
We found that 65\% chart-caption pairs made by professional authors match emphases (blue {\color[RGB]{140,158,212}$\blacksquare$} segment in the top bar of Figure~\ref{fig:survey-results}); the chart emphasizes the author's message in the text, and the text explains the visually prominent features in the chart.
Yet, emphasis mismatches occur regularly; in 35\% of the chart-caption pairs, the chart emphasizes a feature in the data different from the text, and the text either describes features that are not visually prominent (26\%; first orange {\color[RGB]{224,164,122}$\blacksquare$} bar in Figure~\ref{fig:survey-results}) or only describes how the chart encodes information (9\%; first gray {\color[RGB]{192,192,192}$\blacksquare$} bar in Figure~\ref{fig:survey-results}).
Such mismatches suggest that there is some room for improvement, even in professionally authored documents.

\vspace{0.05in}
\noindent\textbf{\textit{Tableau Public authors often write basic captions.}}
In Tableau Public, the overwhelming majority (93\%) of the text descriptions for charts are basic captions that do not describe the specific features in the charts (gray {\color[RGB]{192,192,192}$\blacksquare$} segment in the bottom  bar of Figure~\ref{fig:survey-results}). 
We hypothesize that this result is due in part because the Tableau Public authoring software defaults to providing  basic captions. Authors do not have to put in the extra effort required to discuss the features visible in the charts. 
Unfortunately, prior work has found that readers find such basic captions of little use~\cite{lundgard2021accessible} and that the basic captions play no role in helping readers understand what they should take away~\cite{kim2021towards}.
Hence, the high ratio of basic captions indicates that these authors could benefit from further support and guidelines for authoring charts and text.
\section{Usage Scenario}
\label{sec:usage-scenario}

\begin{figure*}
    \centering
    \begin{subfigure}[b]{0.245\textwidth}
        \centering
        \includegraphics[width=\textwidth]{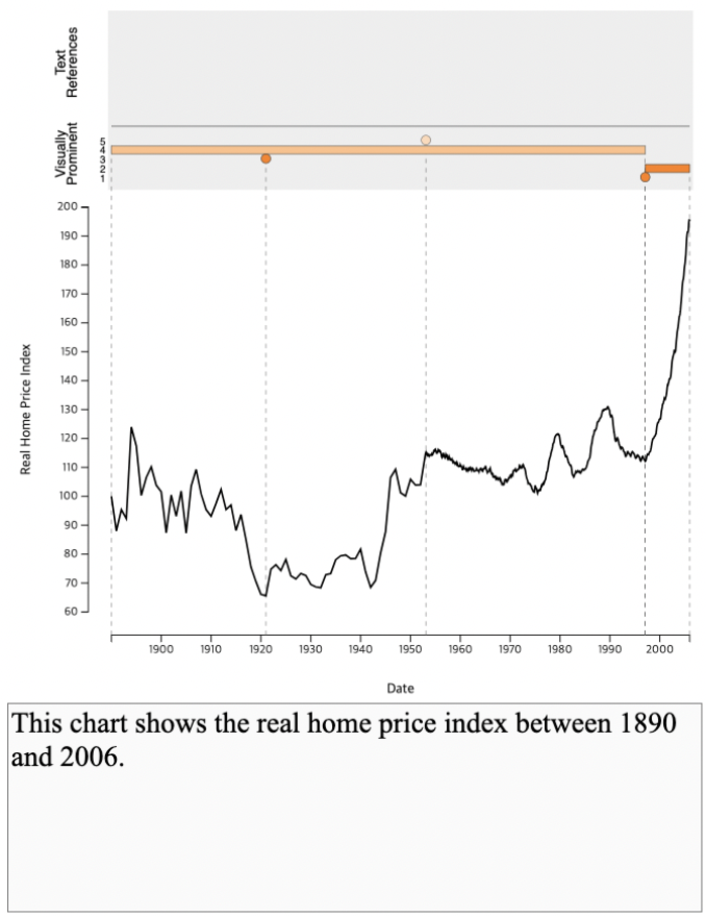}
        \caption{Prominent features \& Basic caption}
    \end{subfigure}
    \begin{subfigure}[b]{0.245\textwidth}
        \centering
        \includegraphics[width=\textwidth]{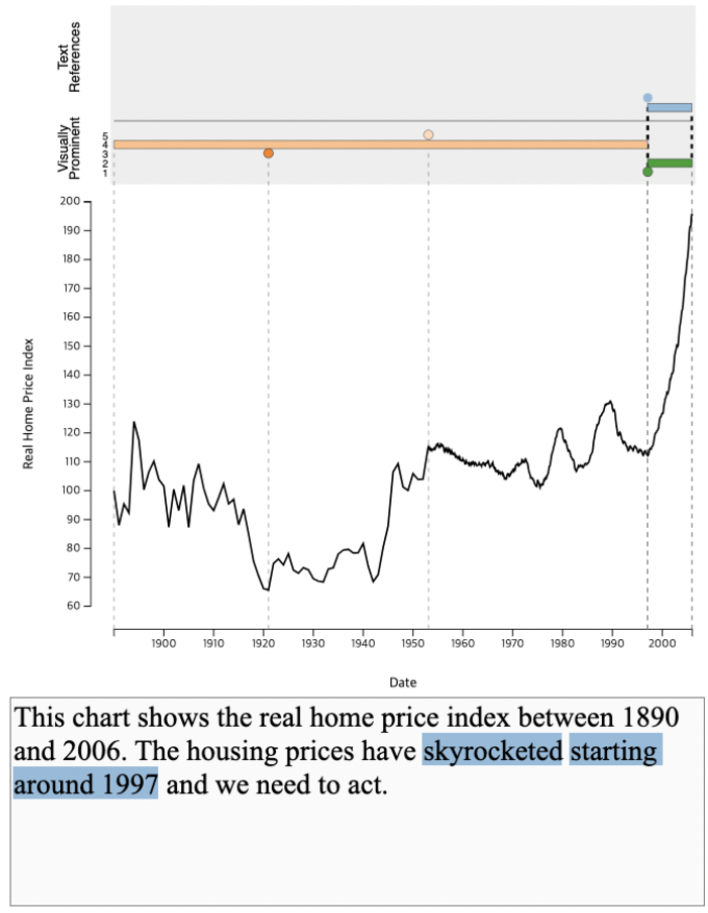}
        \caption{Caption text about prominent feature}
    \end{subfigure}
    \begin{subfigure}[b]{0.245\textwidth}
        \centering
        \includegraphics[width=\textwidth]{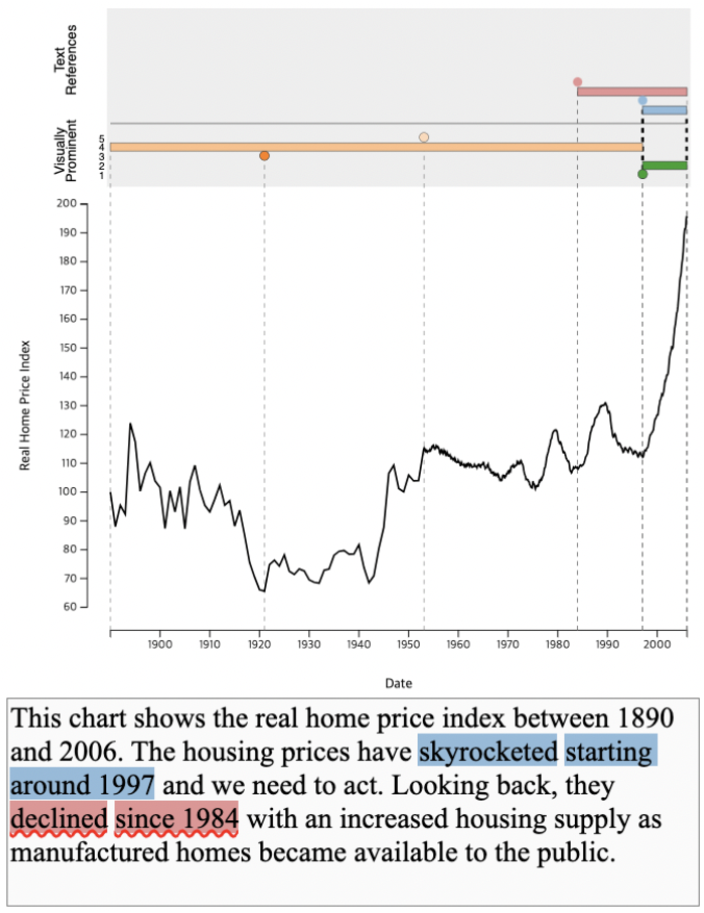}
        \caption{Caption including false information}
    \end{subfigure}
    \begin{subfigure}[b]{0.245\textwidth}
        \centering
        \includegraphics[width=\textwidth]{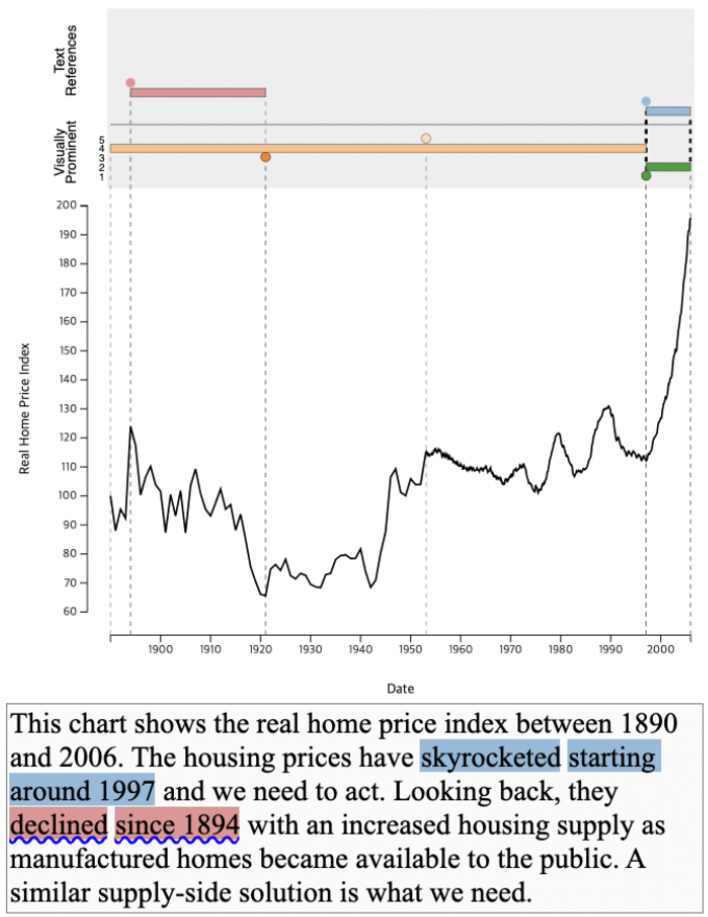}
        \caption{Caption about less prominent feature}
    \end{subfigure}
    \vspace{-0.1in}
    \caption{Views of the \toolname{} interface from the usage scenario. The chart shows the real home price index between 1890 and 2006. (a) Prominent features are shown on top with a basic caption not describing any specific feature. (b) Caption text matches the most prominent visual feature (sharp rise on the right; blue {\color[RGB]{153,185,211}$\blacksquare$} highlight in the UI).
    (c) Typo in the caption text indicated by a red squiggly underline (\includegraphics[height=3px]{figure/red-squiggle}) on `\textit{declined since 1984}'. (d) Caption text matching a less prominent feature, indicated by a blue squiggly underline (\includegraphics[height=3px]{figure/blue-squiggle}) on `\textit{declined since 1894}.'
    }
    \label{fig:user-interface}
\end{figure*}

We first describe a usage scenario of \toolname{} to motivate the design of the tool and illustrate how it guides the user when writing a text caption for a chart.
Tess, a policymaker, wants to add a chart showing the real home price index data to a presentation she plans to give to her fellow policymakers. As a part of her presentation, she would like to make a case for building more homes based on past data. To ensure that she effectively gets her message across, she decides to use the \toolname{} tool.

\vspace{0.05in}
\noindent\textit{\textbf{Viewing visually prominent features in the chart.}} (Figure~\ref{fig:user-interface}a above textbox) Tess starts by loading the data into \toolname{}. The time-series chart shows the real home price index over time. The tool displays the visually prominent features in the chart as orange circles ({\color[RGB]{223,123,51}$\CIRCLE$}) and bars ({\scalebox{1.5}[1]{\color[RGB]{223,123,51}$\blacksquare$}}) just above the chart. 
\toolname{} show  the recent low point in 1997 and the rising segment afterward as the top two most prominent features (darker shade of orange {\color[RGB]{223,123,51}$\blacksquare$}), followed by the global minimum point around 1920 and two other less prominent features (lighter shades of orange {\color[RGB]{238,189,141}$\blacksquare$}).

\vspace{0.05in}
\noindent\textit{\textbf{Typing a basic caption.}} (Figure~\ref{fig:user-interface}a) Seeing the time range of the chart, she types \textit{``The chart shows the real home price index between 1890 and 2006.''} in the text box below the chart and hits the \textsc{[Shift-Enter]} command to run text analysis and assess text emphasis with respect to the chart.  
After the code runs to completion and the textbox is re-enabled, she sees that there is no change in the region above the chart.
Based on this information, she understands that the text she typed in has no reference to any specific features in the chart.

\vspace{0.05in}
\noindent\textit{\textbf{Typing caption text that matches the most prominent visual feature.}} (Figure~\ref{fig:user-interface}b) Tess then returns to thinking about the most prominent feature. 
She hovers over the orange circle ({\color[RGB]{223,123,51}$\CIRCLE$}) closest to the bottom and the orange bar ({\scalebox{1.5}[1]{\color[RGB]{223,123,51}$\blacksquare$}}) right above it to see where they lie on the chart.
She observes the spike after 1997 in the context of the chart and convinces herself that the spike is not only visually prominent but also that she can point to this spike as proof of the dire situation of the housing market.
She decides to cover the feature in her caption and types
\textit{``The housing prices have skyrocketed starting around 1997 and we need to act.''} After hitting \textsc{[Shift-Enter]} to run analysis on the text, the interface highlights the phrase \textit{`skyrocketed starting around 1997'} in blue {\color[RGB]{153,185,211}$\blacksquare$} and displays a blue circle ({\color[RGB]{153,185,211}$\CIRCLE$}) on the year 1997, the endpoint explicitly mentioned in the text, and a bar ({\scalebox{1.5}[1]{\color[RGB]{153,185,211}$\blacksquare$}}) starting in the year 1997 to show the reference between the chart and caption.
The interface also highlights the top two prominent chart features in green {\color[RGB]{97,158,58}$\blacksquare$} to indicate that Tess has matched the emphasis in the caption she has written so far.

\vspace{0.05in}
\noindent\textit{\textbf{Typing caption text with an error.}} (Figure~\ref{fig:user-interface}c)
Tess now looks for falling trends in the chart and discovers one between 1894 and 1921. She looks up the reason for the fall and thinks that it would support her message well; she types, \textit{``Looking back, they declined since \textbf{1984} with an increased housing supply as manufactured homes became available to the public.''}
Tess hits \textsc{[Shift-Enter]} and this time, she is surprised to see a red squiggly underline 
(\includegraphics[height=3px]{figure/red-squiggle})
under the phrase \textit{`declined since 1984'} in her caption. When she hovers over the text \textit{`declined since 1984'}, she sees that the time segment that she is referring to (red bar ({\scalebox{1.5}[1]{\color[RGB]{207,145,147}$\blacksquare$}}) above the chart in Figure~\ref{fig:user-interface}c) is not the one she intended and soon realizes that she mistyped \textit{`1984'} instead of \textit{`1894'}. 

\begin{figure}
    \centering
    \vspace{-0.2in}
    \includegraphics[width=\linewidth]{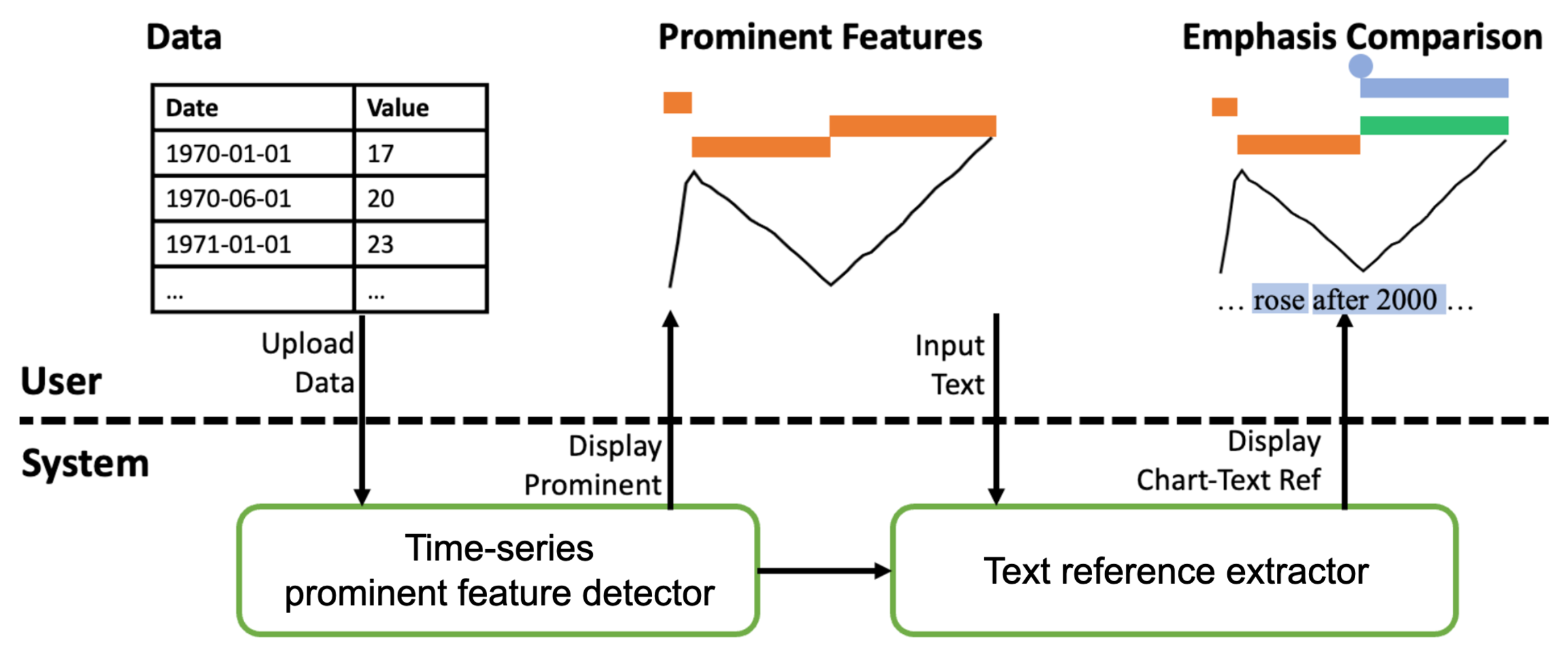}
    \vspace{-0.2in}
    \caption{\toolname{}~tool overview. A user first uploads time-series data (including axis ranges and aspect ratio). Based on the input, the \textit{time-series prominent feature detector} detects the visually prominent features and displays them to the user (Section~\ref{sec:prominent-detector}). The user can type text based on the prominent features they see in the chart. When triggered, the \textit{text reference extractor} identifies the references between the chart and text and displays them to the user, with comparisons with the prominent features (Section~\ref{sec:reference-identifier}).}
    \label{fig:system}
\end{figure}

\vspace{0.05in}
\noindent\textit{\textbf{Pushing through with caption text about a less prominent feature.
}} (Figure~\ref{fig:user-interface}d)
Tess fixes the typo and completes her caption by adding the sentence \textit{``A similar supply-side solution is what we need.''} 
She then confirms the change by pressing \textsc{[Shift-Enter]}. 
She finds that the detected time range has been revised to the one she initially intended, but she realizes that there is a blue squiggly underline (\includegraphics[height=3px]{figure/blue-squiggle})
under the same phrase \textit{`declined since 1894'}.
She sees that the red bars ({\scalebox{1.5}[1]{\color[RGB]{207,145,147}$\blacksquare$}}) matched to the phrase do not match with any of the top five prominent features but decides that she wants to push forward with her caption.
Before finalizing her caption, she looks over the unmatched prominent features, still shown in orange ({\color[RGB]{223,123,51}$\CIRCLE$}), and thinks about whether there are additional features she should describe in her caption.
She sees that the third most prominent feature corresponding to the global minimum in 1921 is the other end point of the downward trend she just wrote about and decides that she does not need to describe the point explicitly.
Finally, Tess looks over the fourth and the fifth most prominent features but decides that they are both irrelevant to her needs and not very prominent based on the lighter shade of orange {\color[RGB]{238,189,141}$\blacksquare$}.
She concludes the chart-caption authoring process.

\begin{figure}[t]
    \centering
    \includegraphics[width=0.85\linewidth]{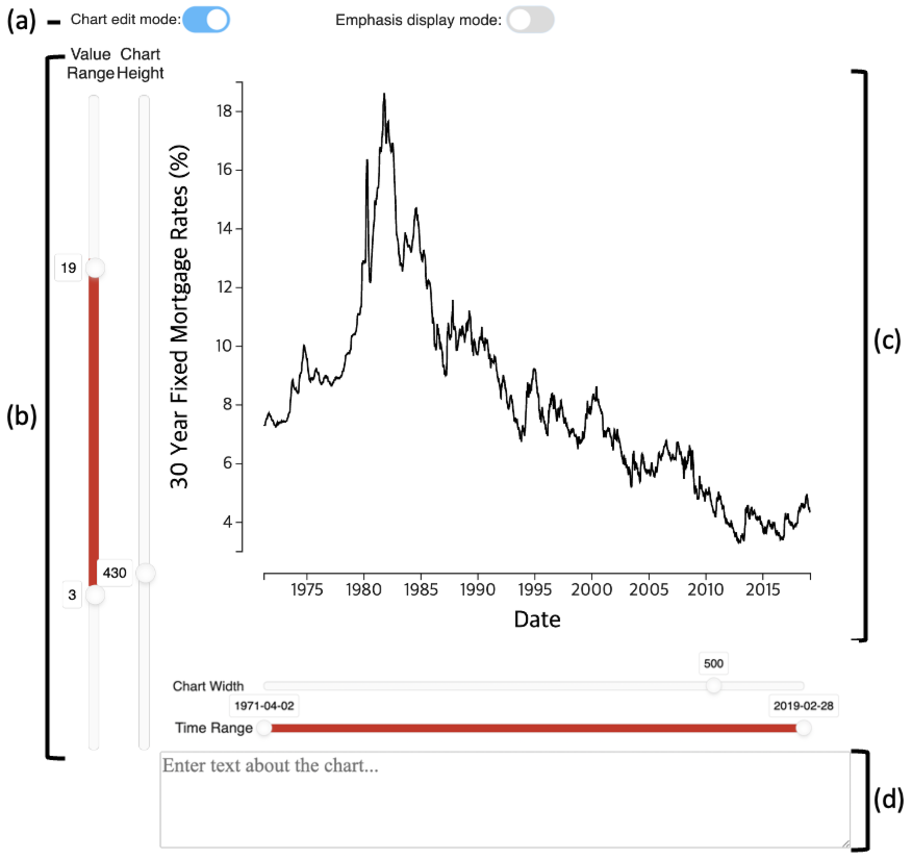}
    \vspace{-0.1in}
    \caption{\toolname{}'s interface 
    for authoring charts and captions. (a) The interface includes two switches, the \textit{chart edit mode} for toggling the sliders (Component (b)) and the \textit{emphasis display mode} that toggles the display of references (Figure~\ref{fig:teaser}a) and the visually prominent features (Figure~\ref{fig:teaser}b). The figure shows the state with chart edit mode on and emphasis display mode off. (b) When the chart edit mode is on, the user can manipulate the vertical and horizontal sliders to edit the dimensions of the chart and the axes ranges. (c) The user can hover over the chart to view tooltips showing the underlying data values. (d) The user can type the caption in the textbox.}
    \label{fig:base-interface}
\end{figure}

\begin{figure*}
    \centering
    \begin{subfigure}{0.42\linewidth}
    \centering
    \includegraphics[width=0.9\linewidth]{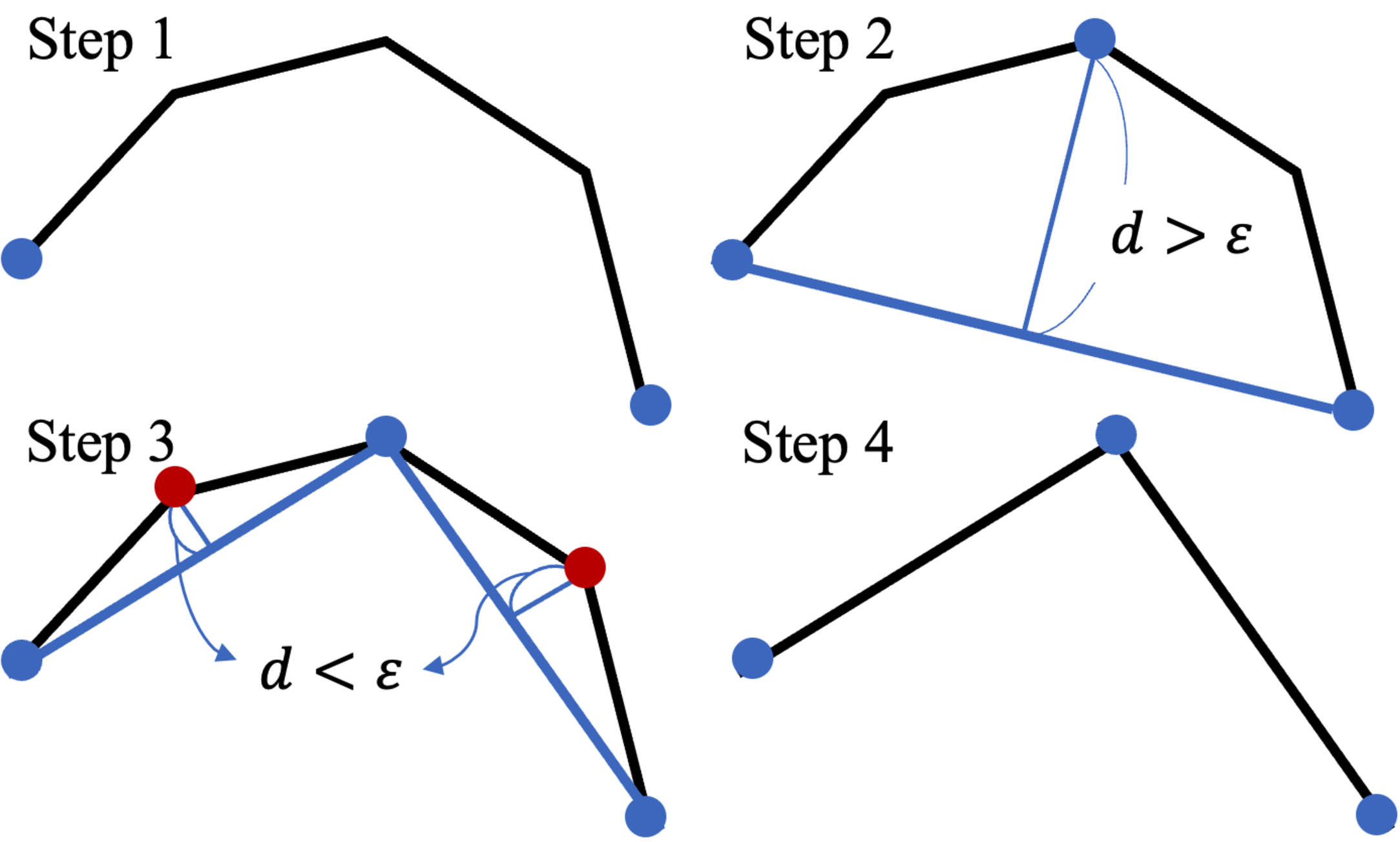}
    \caption{Steps of the RDP algorithm}
    \label{fig:rdp-alg}
  \end{subfigure}
  \hfill
  \begin{subfigure}{0.52\linewidth}
    \centering
    \includegraphics[width=0.9\linewidth]{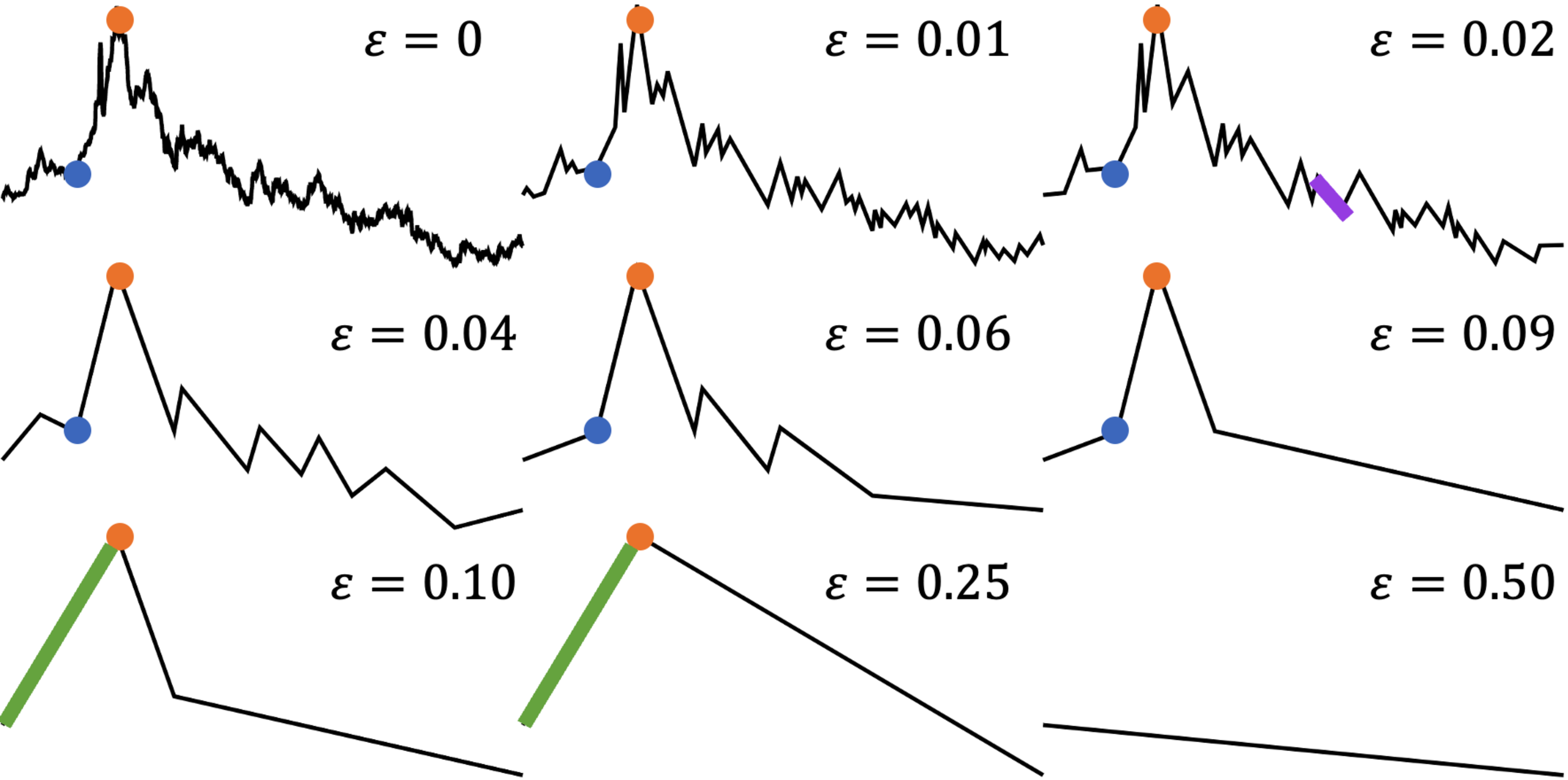}
    \caption{Results of the RDP algorithm at various $\varepsilon$ values (diagonal = 1)}
    \label{fig:epsilon}
  \end{subfigure}
    \caption{(a) [Step 1] The RDP algorithm~\cite{ramer1972iterative,douglas1973algorithms} takes an input polyline and a distance threshold parameter $\varepsilon$. [Step 2] It first checks whether the distance between the line segment connecting its endpoints and the farthest point on the polyline ($d$) is greater than $\varepsilon$. If so, the farthest point captures an important deviation in the polyline. [Step 3] The algorithm breaks the polyline at this point and is recursively applied to both halves. If $d < \varepsilon$, the deviation at the farthest point is considered not important. [Step 4] The algorithm simplifies the polyline to the line segment between the endpoints, resulting in a simplified polyline.
    (b) Applying the RDP algorithm on the time series in Figure~\ref{fig:user-interface} with various $\varepsilon$ values yields simplified curves with various levels of detail; lower values of $\varepsilon$ preserves local details whereas higher values of $\varepsilon$ draws out the global patterns. For very high values of $\varepsilon$, the polyline degenerates into a single line segment ($\varepsilon = 0.50$).
    The most prominent orange peak ({\color[RGB]{207,113,52}$\CIRCLE$}) in this chart persists through all $\varepsilon$ values between $0$ and $0.25$ and hence has an $\varepsilon$-persistence of $0.25$. In comparison, the blue point ({\color[RGB]{77,103,184}$\CIRCLE$}) to the left of the peak persists through $\varepsilon$ values between $0$ and $0.09$ and hence has an $\varepsilon$-persistence of $0.09$.
    The green rise up to the peak ({\scalebox{1.5}[1]{\color[RGB]{122,161,71}$\blacksquare$}}), the second most prominent feature in this chart, is first seen at $\varepsilon = 0.10$ and persists until $\varepsilon = 0.25$ and hence has an $\varepsilon$-persistence of $0.15$. On the other hand, the less prominent trend along the decrease (purple; {\scalebox{1.5}[1]{\color[RGB]{130,62,220}$\blacksquare$}}) persists only briefly around the $\varepsilon$ value of $0.02$ and hence has negligible $\varepsilon$-persistence. 
    }
    \label{fig:rdp}
\end{figure*}

\section{Components of \toolname{}}
\toolname{} allows the user to edit both the chart and caption (Figure~\ref{fig:base-interface}).
The user can not only write captions (Figure~\ref{fig:base-interface}d) but also edit the dimensions and the x- and y-axis ranges (Figure~\ref{fig:base-interface}b). The tool provides two main features in addition to this basic chart and caption editing interface (Figure~\ref{fig:system}):
(1) The \textit{time-series prominent feature detector} computes the visual prominence of the chart features in the time-series line chart (Figure~\ref{fig:teaser}b) and
(2) the \textit{text reference extractor} identifies references between the caption text and the time-series line chart (Figure~\ref{fig:teaser}a).

\subsection{Time-Series Prominent Feature Detector}
\label{sec:prominent-detector}

The \textit{time-series prominent-feature detector} identifies visually prominent features of a given time-series line chart by looking for features that persist through multiple levels of detail.

Prior work~\cite{jugel2014m4,keogh2001dimensionality,rong2017asap,Rosen2020LineSmoothAA} has sought ways to simplify line charts to enhance the perceivability of patterns in line charts that are often obscured by the variations in the chart.
However, these methods leave the interpretation of the features to the users and do not provide ways to compare the features.
Based on the strength of the Ramer-Douglas-Peuker (RDP) line simplification algorithm (Figure~\ref{fig:rdp}a) in enhancing important extrema in time series~\cite{Rosen2020LineSmoothAA}, we modify the vanilla RDP algorithm to compute \textit{$\varepsilon$-persistence} for measuring visual prominence of both point and trend features.
 
The RDP algorithm (Figure~\ref{fig:rdp}a) takes a polyline and a distance threshold parameter $\varepsilon$, and modifying $\varepsilon$ changes the level of detail in the resulting simplified polyline (Figure~\ref{fig:rdp}b).
Our key observation is that visually prominent features persist through multiple $\varepsilon$ values (e.g., orange peak ({\color[RGB]{207,113,52}$\CIRCLE$}) in Figure~\ref{fig:rdp}b).

Based on the observation, our persistence algorithm initiates by running 
the RDP algorithm on the input time series data multiple times. On each run, the algorithm varies the $\varepsilon$ threshold, stepping through the range of values $[0.0, 0.25]$ with a step size of $0.01$, 
relative to a chart whose diagonal length is normalized to equal
$1$.
This normalization allows us to reduce the dependencies of the algorithm on the scale of the rendered charts (e.g., the algorithm would treat a 300px-by-400px chart the same as a 600px-by-800px chart).
We chose the step size to have a sufficient amount of resolution into the range of $\varepsilon$ values while being able to be run in real-time.
We do not proceed to higher values of $\varepsilon$ as most charts degenerate into just two points and no longer provide useful signal (Figure~\ref{fig:rdp}b, $\varepsilon = 0.50$).

Finally, we 
define the $\epsilon$-persistence of each point in the input time series as the
greatest $\varepsilon$ value at which the point is deemed important and included in the simplified polyline by the RDP algorithm,
capped to 0.25 by our choice of the range of $\varepsilon$ values.
In Figure~\ref{fig:rdp}b, for example, the prominent orange peak ({\color[RGB]{207,113,52}$\CIRCLE$}) persists all the way up to $\varepsilon = 0.25$ and therefore has an $\varepsilon$-persistence of $0.25$.

We extend this $\varepsilon$-persistence measure for points to a measure of $\varepsilon$-persistence of trends between any two points $p_i$ and $p_j$ on the original polyline.
Observe the green trend ({\scalebox{1.5}[1]{\color[RGB]{122,161,71}$\blacksquare$}}) to the left of the orange peak ({\color[RGB]{207,113,52}$\CIRCLE$}) in Figure~\ref{fig:rdp}b.
Below $\varepsilon = 0.10$, the trend is broken into smaller trends by the blue point ({\color[RGB]{77,103,184}$\CIRCLE$}) left of the orange peak ({\color[RGB]{207,113,52}$\CIRCLE$}).
Somewhere above $\varepsilon = 0.25$, the orange peak ({\color[RGB]{207,113,52}$\CIRCLE$}) is removed and the trend is absorbed into a larger trend.
Thus, the green trend ({\scalebox{1.5}[1]{\color[RGB]{122,161,71}$\blacksquare$}}) persists from $\varepsilon = 0.10$ to $\varepsilon = 0.25$ and hence has $\varepsilon$-persistence of $0.15$.
Generalizing this observation, we can compute $\varepsilon$-persistence of trends between any two points $p_i$ and $p_j$ by subtracting the maximum $\varepsilon$-persistence of points lying between $p_i$ and $p_j$ from the minimum $\varepsilon$-persistence for $p_i$ and $p_j$ and adding 0.01, the $\varepsilon$ step size we use.

\toolname{} displays the top five most prominent features above the chart in order (Figure~\ref{fig:teaser}b), the most prominent closest to the chart, using circles to depict point features and bars to depict trend features.
To help users get a sense of the degree of prominence of these features, the tool displays more prominent features using darker shades ({\color[RGB]{207,113,52}$\blacksquare$}/{\color[RGB]{97,158,58}$\blacksquare$}; e.g., the two most prominent features in Figure~\ref{fig:user-interface}a) and less prominent features using lighter shades ({\color[RGB]{245,216,183}$\blacksquare$}/{\color[RGB]{168,205,151}$\blacksquare$}; e.g., the fifth most prominent feature in Figure~\ref{fig:user-interface}a).

\subsection{Text References Extractor}
\label{sec:reference-identifier}

\begin{figure}
    \centering
    \includegraphics[width=\linewidth]{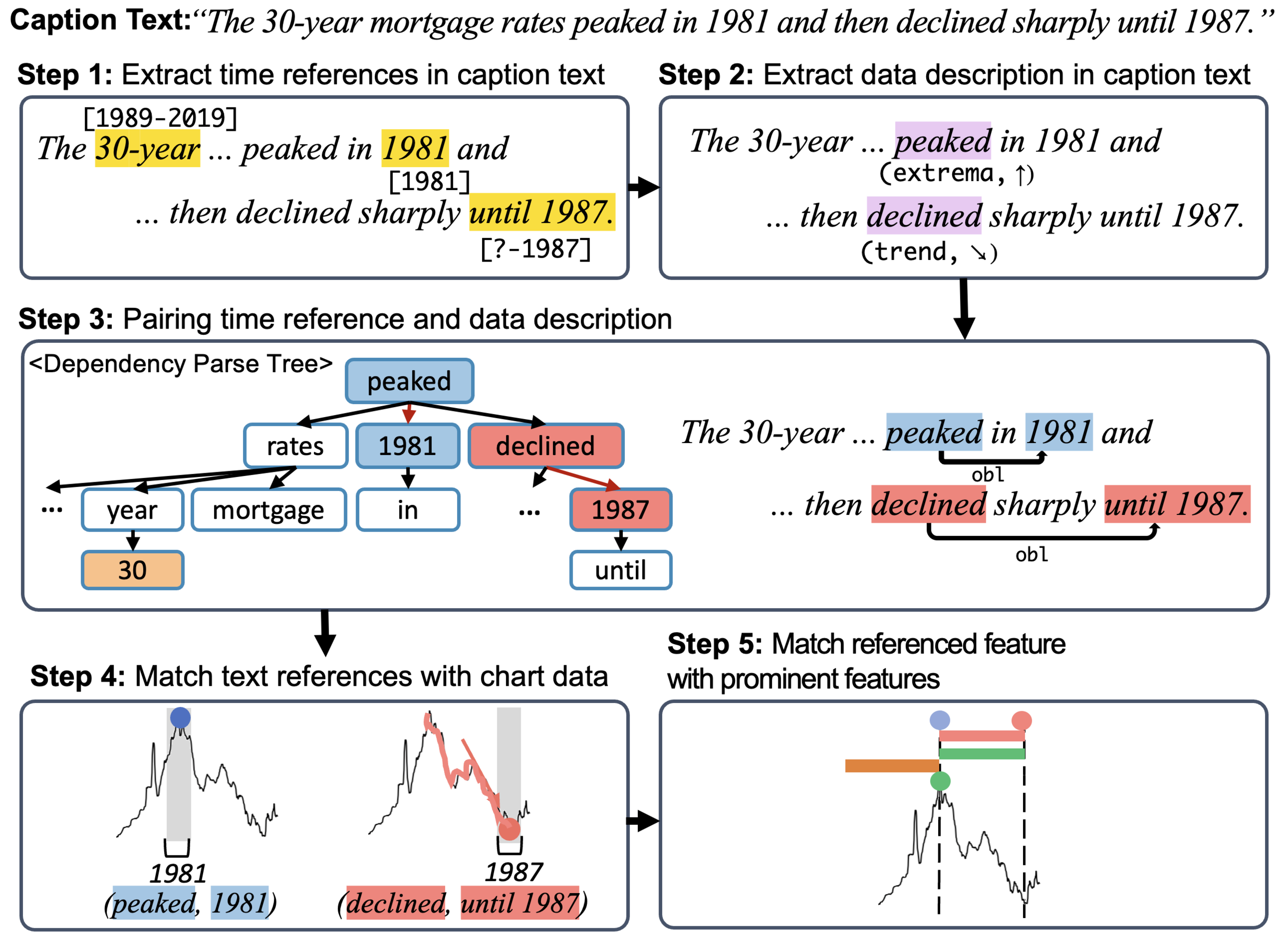}
    \caption{The five steps of the \textit{text references extractor}. 
    }
    \vspace{-0.15in}
    \label{fig:pipeline}
\end{figure}

Our \textit{text reference extractor} uses a five-step pipeline to determine the matches between chart features and caption text (Figure~\ref{fig:pipeline}). 

\vspace{0.05in}
\noindent \textbf{\textit{Step 1. Extract time references in caption text.}}
To detect the time ranges described in the caption text, our tool first uses the Named Entity Module in the Stanford CoreNLP toolkit~\cite{chang2012sutime,finkel2005incorporating,manning2014stanford}. It first identifies phrases describing points in time (e.g., \textit{`1970'}, \textit{`March 2020'}) or durations (e.g., \textit{`the last six months'}).
Since the time-series data in the chart is often represented at a finer granularity than the dates and durations mentioned in the caption text, we convert the detected time points and durations into a set of time points at the finest granularity of the time-series data. In practice, the time-series data we have worked with is at a granularity of days, weeks, months or years. 

The time points mentioned in the caption text occasionally signify the start or the end point of a time range.
For example, \textit{`between 1970 and 1980'} or \textit{`from 1970 to 1980'} indicates the start point 1970 and 1980.
To detect these start and end points, we utilize the context template patterns near the time words (Table~\ref{tab:boundary-points}).
In the example phrases, our tool would detect the time range 1970-1980.
However, caption text sometimes gives only one endpoint of a time range (e.g., `\textit{since Nov 1997}', `\textit{after March 2020}').
In these cases, we detect one endpoint and leave the other endpoint undetermined (e.g., 1997/11-?, 2020/03-?).

\vspace{0.05in}
\noindent \textbf{\textit{Step 2. Extract data descriptions in caption text.}} 
We next extract trend descriptions (i.e., upward, downward) and local extrema (i.e., maximum and minimum) in the caption text by looking for a predefined set of keywords (Table~\ref{tab:feature-words}) in the caption.
We compiled these keywords by examining 
captions collected through our survey of line charts in the wild
and expanding them using a thesaurus~\cite{merriam2022thesaurus}.

We first compared lemmas (i.e., base forms of words; e.g., `\textit{rising}' - `\textit{rise},' `\textit{soared}' - `\textit{soar}') of each word in the input sentence with the lemmas of words in our compiled list of keywords to find any exact matches.
To capture any synonyms that we may have missed in our list, we use the cosine similarity of the BERT contextual embedding vectors~\cite{devlin-etal-2019-bert} between the words in the input sentence and the words in the keywords list.
We obtained the BERT contextual embedding vectors of the keywords by running a pre-trained BERT model~\cite{devlin-etal-2019-bert} on the sentences from which the words originated and extracting the results of the final layer.
For each of the words in the input sentence, we similarly obtain its BERT contextual embedding vector. We consider the word in the sentence a description of the data if the cosine similarity of its BERT contextual embedding vector and any of the BERT contextual embedding vectors of the keywords is greater than the empirically determined threshold of $0.7$.

\vspace{0.05in}
\noindent \textbf{\textit{Step 3. Pairing time references with data descriptions.}} 
Complex caption sentences may sometimes refer to more than one point or duration in time along with more than one time feature. 
For example, the sentence \textit{``The 30-year fixed mortgage rates peaked in 1981 and then declined sharply until 1987''} includes the time duration \textit{`30-year'}, and points in time \textit{`1981'} and \textit{`until 1987,'} with features \textit{`peaked'} and \textit{`declined.'}

To connect the times with their features, we first use the Stanford CoreNLP dependency parser~\cite{chen2014fast,manning2014stanford} to obtain a dependency tree for the caption (Figure~\ref{fig:pipeline} Step 3 left). 
We match each of the time references to the closest data description within the dependency tree that does not have a closer time reference.
When two time references complement each other, with one only mentioning the start point and the other only mentioning the end point,
we combine them into a single time range.
For example, in Figure~\ref{fig:results}b, Sentence 2, 
\textit{``From 1950, North Korea's GDP increased quite rapidly until 1985,''}
the data description \textit{`increased'} is closest to the two time references, \textit{`from 1950'} (1950-?) and \textit{`until 1985'} (?-1985) and the tool combines the two end dates into a single range, 1950-1985.
We discard time references or data descriptions not matched through this process.
In our example, we follow the
relation from \textit{`1981'} and \textit{`1987,'} to pair the corresponding time references with the data descriptions \textit{`peaked'} and \textit{`declined'} at a distance of $1$, respectively. 
On the other hand, the time phrase, \textit{`30-year'} remains unmatched and is therefore discarded.

\begin{table}
    \centering
    \caption{List of context template patterns near the mention of time \textit{T} that we use to determine whether \textit{T} is a start point or an end point of a time range.}
    \includegraphics[width=\linewidth]{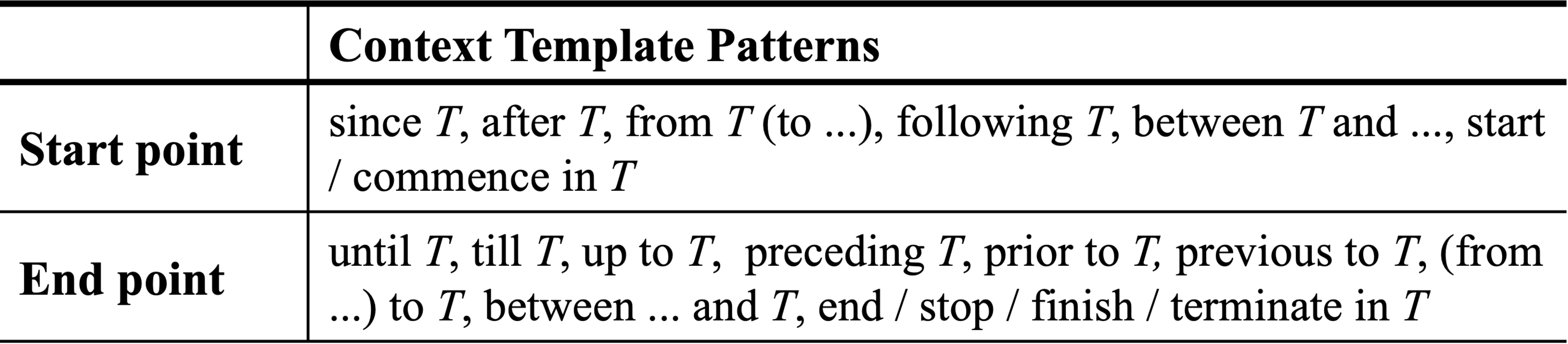}
    \label{tab:boundary-points}
\end{table}

\begin{table}
    \centering
    \caption{The word list \toolname{} uses to identify data descriptions in caption text.}
    \includegraphics[width=\linewidth]{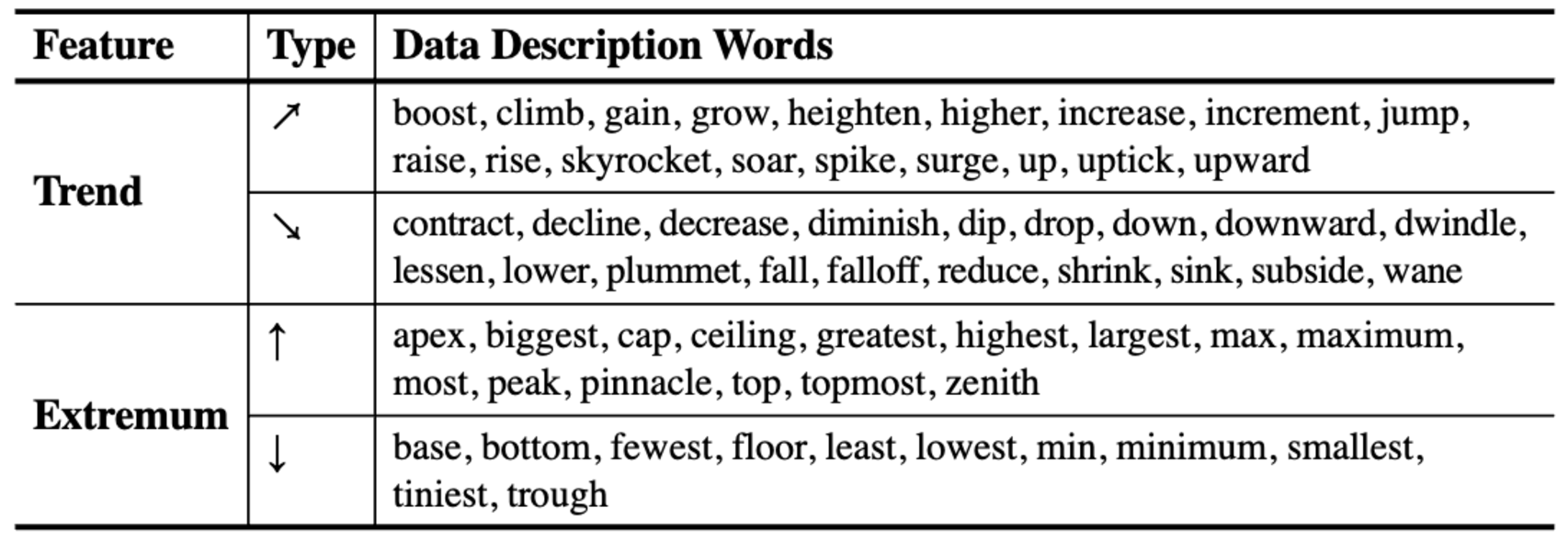}
    \vspace{-0.15in}
    \label{tab:feature-words}
\end{table}

\vspace{0.05in}
\noindent \textbf{\textit{Step 4. Match text references with chart data.}}
The data descriptions and time references in the caption text obtained from the previous three steps may require further disambiguation to accurately pinpoint the referenced features in the chart. 
For example, the time reference \textit{`1981'} could refer to any of the points in the chart whose year is 1981. The time reference \textit{`until 1987'} could end at any of the points in the chart whose year is 1987, and the start point is yet to be determined.

To find a suitable selection, we join the information specified in the data descriptions and the time references with the chart data.
When the data description refers to a local maximum, our tool infers that the chart point being referenced is the point within the time reference with the greatest value; if it refers to a local minimum, our tool infers that the chart point being referenced is the point within the time reference with the smallest value.
Hence, in our example, we select the global maximum point in the chart for the pair (\textit{`peaked'}, \textit{`1981'}) (blue point ({\color[RGB]{152,167,214}$\CIRCLE$}) in Figure~\ref{fig:pipeline}).
For rising trends, we select the point with the minimum value among the start point candidates and the point with the maximum value among the endpoint candidates, and vice versa for falling trends.
Thus, for the pair (\textit{`decreased'}, \textit{`until 1987'}), the tool matches the decreasing trend between the maximum point in 1981 and the minimum point in 1987 (green line ({\scalebox{1.5}[1]{\color[RGB]{132,188,121}$\blacksquare$}}) in Figure~\ref{fig:pipeline}).

\toolname{} highlights the time references and data descriptions in the caption text and adds bars (for trend features) and circles (for point features) in the region above the prominent features (Figure~\ref{fig:teaser}a) using the same colors. Users can hover over either the text highlighting or the bars and circles to view the set of referenced points on the chart.
During this process, \toolname{} performs a basic check for factual errors between references to trend features in the text and the actual change of data between the two endpoints of the trend feature. If the data reference is an upward trend but the data decreases, or vice versa, our system detects a factual error. For example, we detect pair (\textit{`soared'}, \textit{`from 1980 to 1991'}) in the first sentence of Figure~\ref{fig:teaser}. Although the text indicates that the value in 1991 would be higher than the value in 1980, this actually is not the case.
The tool alerts authors of such factual errors by drawing a red squiggly underline (\includegraphics[height=3px]{figure/red-squiggle}) on the time references and data descriptions in the text, similar to spell-checkers.

\vspace{0.05in}
\noindent \textbf{\textit{Step 5. Match referenced feature with prominent features.}}
We finally compare the references with the prominent chart features returned by our prominent feature detector.
We consider points features as a match if they are exactly the same.
On the other hand, we 
detect a match between text references to time segments and a visually prominent trend if the intersection of the points in the two sets covers at least 95\% of the union of the two sets.
This way, we are able to detect that (\textit{`peaked'}, \textit{`1981'}) matches the most prominent feature and that trend feature referred to by (\textit{`decreased'}, \textit{`until 1987'}) matches the fourth most prominent feature, whereas its endpoint of in 1987 does not match any feature as it is off from the third most prominent feature by couple weeks.

If a prominent feature is matched to a reference, \toolname{} highlights the feature in green {\color[RGB]{97,158,58}$\blacksquare$} (e.g., Figure~\ref{fig:teaser} most prominent and fourth most prominent features).
On the contrary, if the feature the caption text refers to is not matched to any text (e.g., Figure~\ref{fig:teaser} Sentence 3), the tool alerts the authors of the emphasis mismatch by drawing a blue squiggly underline (\includegraphics[height=3px]{figure/blue-squiggle}) on the time references and data descriptions in the caption task, similarly to grammatical errors in grammar-checkers.
\section{Results and Evaluation}

\begin{figure*}
    \centering
    \begin{subfigure}[b]{0.245\textwidth}
        \centering
        \includegraphics[width=\textwidth]{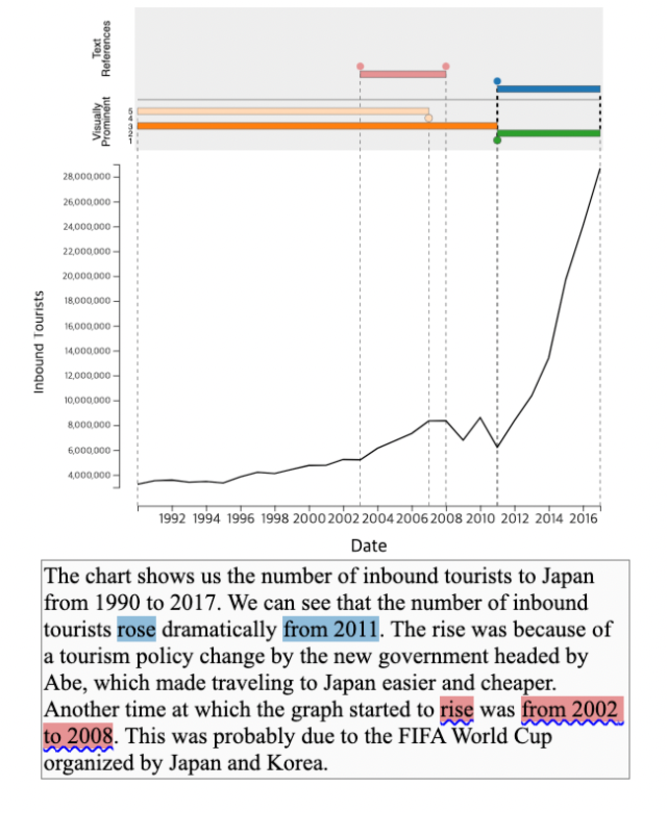}
        \caption{Tourists inbound to Japan}
    \end{subfigure}
    \begin{subfigure}[b]{0.245\textwidth}
        \centering
        \includegraphics[width=\textwidth]{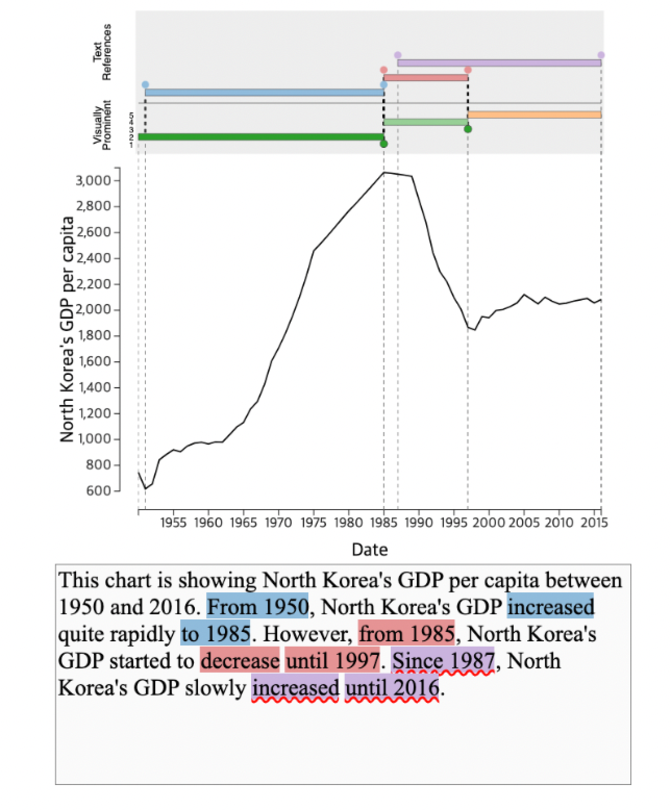}
        \caption{North Korea's GDP per capita}
    \end{subfigure}
    \begin{subfigure}[b]{0.245\textwidth}
        \centering
        \includegraphics[width=\textwidth]{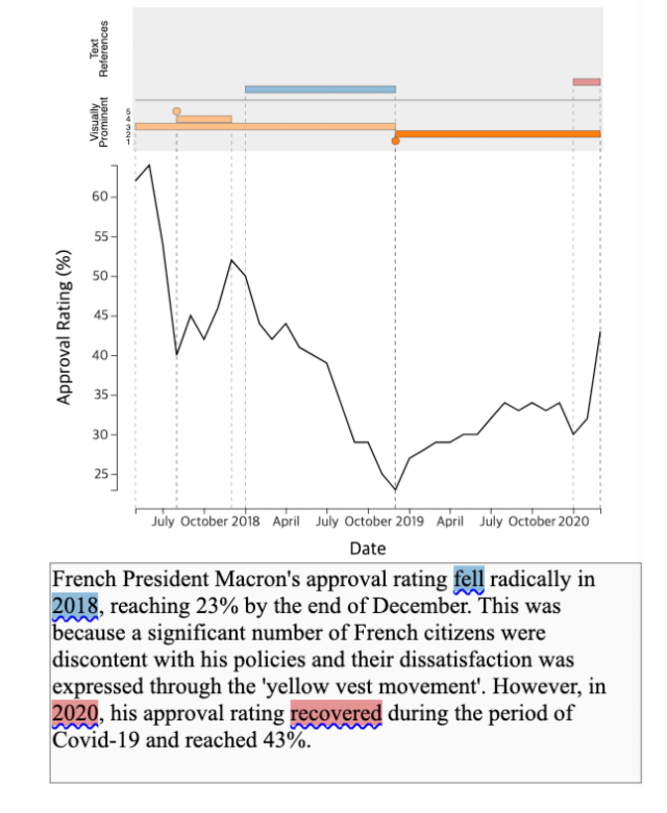}
        \caption{Macron's approval rating}
    \end{subfigure}
    \begin{subfigure}[b]{0.245\textwidth}
        \centering
        \includegraphics[width=\textwidth]{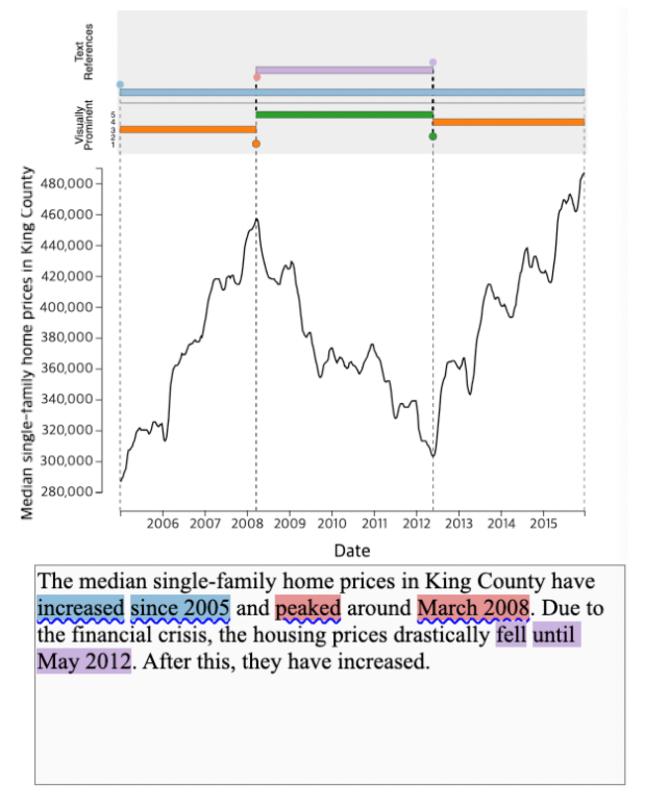}
        \caption{Home prices in King County}
    \end{subfigure}
    \caption{Results of running \toolname{} on charts and captions written by participants in the user study (Section~\ref{sec:user-study}). Participants wrote the chart-caption pairs in (a) and (b) with the \toolname{} tool and the chart-caption pairs in (c) and (d) with the baseline tool. 
    (a), (b), and (c) show chart-caption pairs for which \toolname{} would provide suitable guidance. (a) Sentence 4 and (c) Sentence 3 mention non-prominent features and are underlined in blue (\includegraphics[height=3px]{figure/blue-squiggle}). (b) Sentence 4 includes a typo (\textit{`1987'} instead of \textit{`1997'}), and the tool underlines the typo in red (\includegraphics[height=3px]{figure/red-squiggle}). 
    However, for (d) Sentence 1, \toolname{} fails to correctly capture the intent in the phrase \textit{`increased since 2015'} and captures a wrong time range and also misses the peak around March 2008 due to an artifact of the RDP algorithm and underlines both blue (\includegraphics[height=3px]{figure/blue-squiggle}) despite being mentions of the most and the third most prominent features. The tool also completely misses references in (d) Sentence 3.
    Minor typos and grammatical errors (verb tense, plural-singular, etc.) have been fixed for illustration.
    }
    \label{fig:results}
\end{figure*}

Figure~\ref{fig:results} shows results of running the \toolname{} tool on various charts and captions.
The tool's components function accurately and provide correct guidance in Figure~\ref{fig:results}a-c, but err in cases shown in Figure~\ref{fig:results}d.
To understand how well each component of \toolname{} performs, we ran an evaluation on the time-series prominent feature detector and the text reference extractor.

\subsection{Evaluation: Time-Series Prominent Feature Detector} 
To evaluate our time-series prominent feature detector, we used Kim et al.'s~\cite{kim2021towards} corpus of 43 synthetic and real-world examples.
We used their crowdsourced prominent features as a gold feature set to compare against.
As another baseline for comparison, we also generated chart features using a state-of-the-art prominent feature detector, Contextifier~\cite{hullman2013contextifier}, which uses the value and the first derivative at each point to compute the visual saliency of each point. 

The features we generated with \toolname{} matched the gold feature set far better than features generated by Contextifier.
On average, $1.47$ ($49\%$) of \toolname's top three prominent features matched the gold features, whereas only $0.81$ ($27\%$) of Contextifier's top three prominent features matched the gold features. 
We further analyzed the variance among crowd workers who labeled the gold feature set. That is, we compared each crowd worker's labeling of the top-three features against the average crowd workers' labels of these features and found a match of $1.72$ ($57\%$) of the time. This suggests that there is 
high variation among what people think of as the top three visually prominent features.

We include further analysis of our prominent feature detection algorithm in the supplemental material.

\subsection{Evaluation: Text Reference Extractor}
To understand how reliably our text reference extractor identifies text references to chart features, we ran our text reference extractor on 24 chart-caption pairs with 82 sentences collected through a user study (Section~\ref{sec:user-study}). 
Two of the authors of this paper independently reviewed each of the captions at sentence level along with the participant's stated intended message to determine the references included in each sentence.
The two authors discussed any mismatches and asked a third-party visualization expert for interventions whenever the conflicts could not be resolved between the two authors.

Based on the labels, we ran our text reference extractor on the captions and classified errors in each sentence as one of the following:
\begin{itemize}[nolistsep,leftmargin=*]
    \item \textit{False negative} (FN): the tool failed to extract an existing reference (e.g., Figure~\ref{fig:results}c Sentence 3).
    \item \textit{False positive} (FP): the tool extracted a reference that is non-existent.
    \item \textit{Intention mismatch} (IM): the tool correctly extracts the reference based on what is said in the sentence but detects a feature that is different from the author's intention (e.g., Figure~\ref{fig:results}d Sentence 1; the endpoint of the range is underspecified in the text, but the participant's intended endpoint is clearly 2008, not the end of the chart).
\end{itemize}
A sentence can include several of these errors and is considered correct if it includes no errors.
For three sentences that included apparent typos (one written with the baseline tool (Figure~\ref{fig:results}a), two written with \toolname{} tool), we considered them correct if the text reference extractor correctly identified the mentioned time range.

Of the 81 sentences, the text reference extractor correctly identified the references in 57 sentences (70\%), included FN errors in 22 sentences (27\%), FP errors in 4 sentences (5\%), and IM errors in 2 sentences (2\%). Four of the sentences included multiple errors and were double-counted into multiple categories.
We include further analysis of the error cases in the supplemental material.
\section{User Study}
\label{sec:user-study}
To evaluate how the \toolname{} tool helps authors
write charts and captions, we conducted a within-subjects study comparing the complete \toolname{} tool (\toolname{} condition) against a baseline system that only includes the chart and text authoring interface in Figure~\ref{fig:base-interface} (baseline condition). 
We consider the following two hypotheses:

\vspace{0.05in}
\noindent\textbf{[H1]} Users will find the features in \toolname{} useful in authoring charts and captions.

\vspace{0.05in}
\noindent\textbf{[H2]} The features in \toolname{} are easy to use.

\vspace{0.05in}
We note that the accuracy of the text reference extractor was lower at 56\% at the time of the user study due to an issue with our implementation of the time reference extraction that was later revised.
\subsection{Participants}

We recruited 12 participants through online communities within KAIST and personal referrals.
We required participants to be capable of reading and writing in English and
we administered a test to ascertain skills such as understanding the chart encoding, reading off values, and recognizing trends and extrema~\cite{kim2021towards}.
In order to model the target population of the tool, we further required that the participants regularly author charts and captions.
All of the participants reported having regular exposure to chart-caption authoring for academic purposes (12 participants, P1-P12), presentations (11 participants), coursework (5 participants), industrial purposes (4 participants), articles (2 participants), social media (2 participants), and personal messages (2 participants).

\subsection{Procedure}
We conducted the study online through Google Meet~\cite{google2023meet} with screen sharing.
We began the study with a pre-survey asking about the participant's background in authoring charts and text as well as their domain of expertise.
We then showed the participant a chart and three types of captions: a basic caption, a caption about prominent chart features, and a caption about non-prominent chart features. 
After reading through the chart-caption pairs, they answered which caption is the most effective and their rationale.
We then provided instructions about the baseline and \toolname{} tools for authoring chart-caption pairs.

During the study, the participant chose two distinct time-series line charts either from the real-world corpus from Kim et al.~\cite{kim2021towards} or one of their own.
The participant authored one chart-caption pair with the baseline tool and another pair with the \toolname{} tool in a counterbalanced order.
After the use of each tool, the participant completed a post-task reflection survey that comprised three parts: (1) the participant's intended messages; (2) a usefulness assessment of the guidance provided by the tool and the expected capability of captions generated with the tool in communicating authors' main messages on 5-point Likert scales and free form comments about the benefits and disadvantages of using the tool; and (3) a usability assessment using the System Usability Scale (SUS)~\cite{brooke1996sus} and free-form comments about the usability of the tool.

After completing the two chart-caption pair authoring tasks, we asked the participant to \textit{compare} their experience with the two tools as a post-survey.
We specifically asked the participants to compare the usefulness and ease of use of the two tools in chart-caption authoring on 5-point scales (i.e., Baseline+2, Baseline+1, Neutral, \toolname{}+1, \toolname{}+2) and asked for rationales for their scores.
We finally asked for any free-form comments on the two tools.

The study lasted 60-75 minutes, and we compensated each participant with an equivalent of 15 USD as a direct deposit. 
We include the survey materials and responses in the supplemental material.

\subsection{Results and Discussion}
\begin{figure}
    \centering
    \begin{subfigure}{0.49\linewidth}
    \centering
    \includegraphics[width=\linewidth]{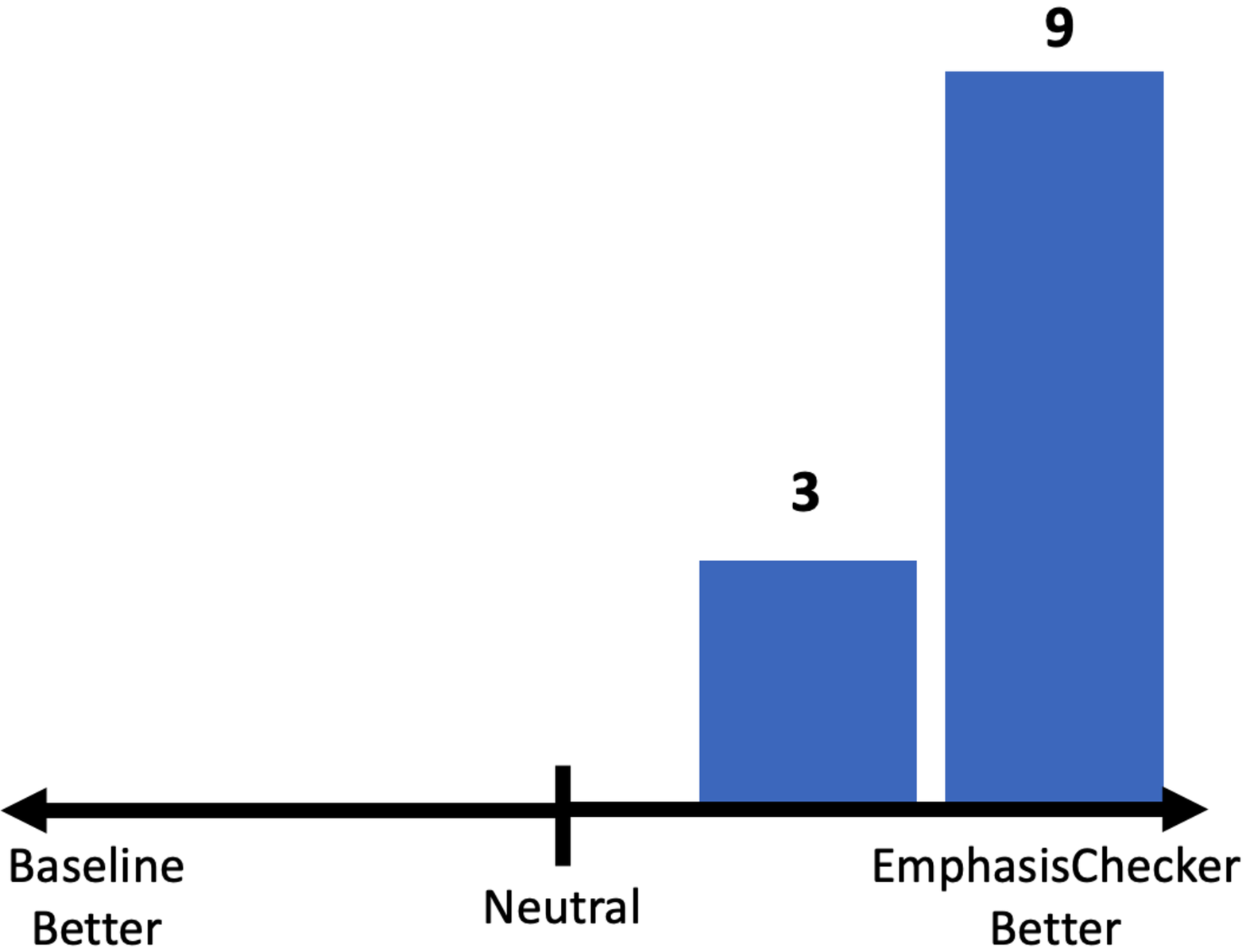}
    \caption{Usefulness}
    \label{fig:usefulness}
  \end{subfigure}
  \hfill
  \begin{subfigure}{0.49\linewidth}
    \centering
    \includegraphics[width=\linewidth]{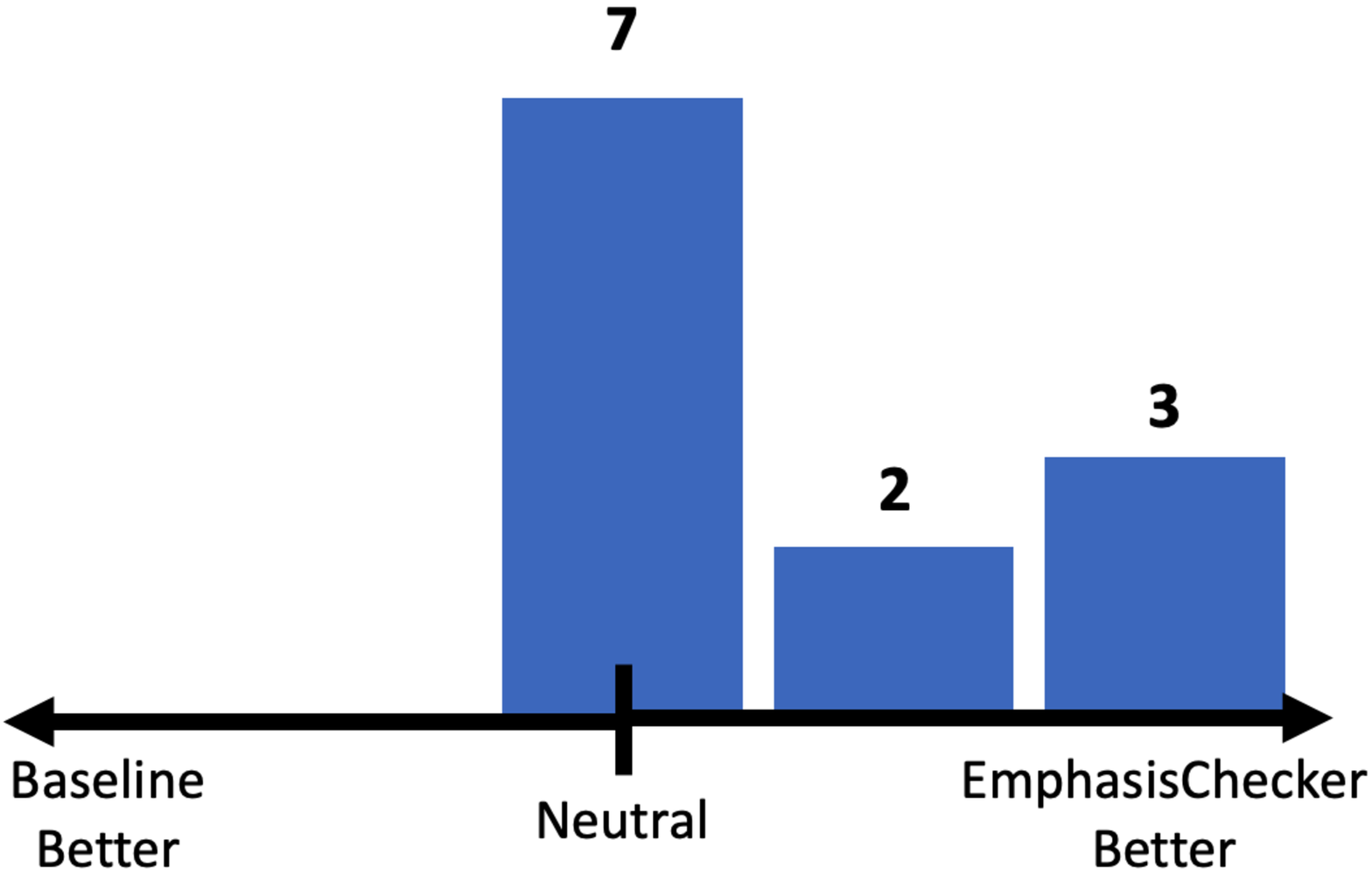}
    \caption{Ease of use}
    \label{fig:ease-of-use}
  \end{subfigure}
    \caption{Participants' comparison of baseline and \toolname{} tools based on their usefulness and ease of use when writing charts and captions. The participants leaned towards \toolname{} for usefulness and were neutral or slightly leaning towards \toolname{} for ease of use.}
    \vspace{-0.15in}
    \label{fig:user-comparisons}
\end{figure}

\vspace{0.05in}
\noindent\textbf{\textit{Assessing H1.}}
All 12 participants stated that \toolname{} is more useful than the baseline tool in authoring (Figure~\ref{fig:user-comparisons}a), with 4 participants explicitly stating their excitement about the technology potentially being applied to their actual authoring routines in their free-form responses.
Comparisons of the ratings of each of the two tools are also in line with their comparisons.
The participants ranked the usefulness of the guides provided by \toolname{} as 4.33 ($\sigma=0.85$), which is significantly higher than the baseline tool's 2.75 ($\sigma=1.09$) (two-sided Wilcoxon signed-rank test; $W(11) = 5.5 < 10$, the critical $W$ value for $\alpha = 0.05$).
The participants also ranked the messaging capability of captions written with \toolname{} as 4.17 ($\sigma=0.69$), also significantly higher than the baseline tool's 3.18 ($\sigma = 0.55$) (two-sided Wilcoxon signed-rank test; $W(8) = 0.0 < 3$, the critical $W$ value for $\alpha = 0.05$).

When we asked the participants about the rationale for their assessment of the usefulness of \toolname{}, 6 of the 12 participants explicitly mentioned that the check for the alignment between the chart and text helped with their authoring process.
For example, P7 wrote, \textit{``(translated) [\toolname{}'s checking feature] gave me a chance to reflect on whether other people would agree on what I described as important in the text.''}

An additional six participants specifically pointed to the prominent feature detection as a component that contributes to the usefulness of \toolname{}.
P8 went further to suggest that the tool could be useful for exploratory data analysis while also precautioning that the features being shown up front in these cases could bias analysis.

All captions we collected through the study described prominent chart features.
We hypothesize that this is because all the participants we had in the pool were not only experienced with authoring charts and captions but also were reminded of the properties of effective chart-caption pairs through a pre-study exercise.

\vspace{0.05in}
\noindent\textbf{\textit{Assessing H2.}} 
While five participants stated that authoring charts and text with \toolname{} was easier,
the majority (7 of 12) of the participants were neutral on the comparison of the ease of use of the two tools (Figure~\ref{fig:user-comparisons}b).
This result is similar to our findings on the SUS scores of each of the two tools; the average SUS score of our tool was 88.33 ($\sigma = 8.06$) was on par with that of the baseline tool at 82.08 ($\sigma = 11.22$) (two-sided paired t-test: $t(11) = 1.96, p = 0.08 > 0.05$).

While it is counterintuitive that \toolname{} with more features was deemed easier to use than the simpler baseline tool by some participants, diving deeper into the participants' comments provides an insight into why this was the case.
Looking first at the comments on the ease of use from the seven participants who were neutral, six of them commented that while \toolname{} includes additional features, the two tools were both very intuitive and easy to use.
The participants who rated \toolname{} as easier to use pointed out that the synergistic effect between the basic authoring interface and the added features reduces the mental burden on the users.
For instance, P4 wrote, \textit{``While rescaling the graph and seeing prominent points, it is easier to focus on the main characteristics of the graph,''} 
pointing to the synergistic effect between the chart-editing interface and the prominent feature display.
P10, who was neutral about the comparison, stated, \textit{``(translated) Physically, Tool A [baseline] was easier because it was simpler but Tool B [\toolname{}] required less effort in authoring charts and was psychologically easier to use.''} 

\vspace{0.05in}
\noindent\textbf{\textit{Participants' reactions to errors in text references.}}
While writing captions with \toolname{}, 10 of the 12 participants experienced errors.
All of the participants immediately noticed errors. 
We note that this is different from how \textit{readers} often fail to see errors in reference extraction methods~\cite{kim2018facilitating}.
We hypothesize that this is because the authors of our tool have a clear intention and expectation of the correct results in mind when they run it, whereas readers do not know the expected results beforehand.
After noticing the error, three of the participants attempted a revision of the text so that their text would be detected correctly, whereas the others read over the text and continued.
In summary, the results suggest that while authors are unlikely to be misled by the system's errors, the errors can lead authors to spend unnecessary efforts in trying to satisfy the system.

However, the prevalence of the errors in the tool occasionally resulted in the failure of participants to realize their error even when \toolname{} behaved correctly.
For example, P12 typed in the wrong year (2018 instead of 2008), which was out of the time range shown in the chart (up to 2015), and \toolname{} correctly identified no match between the text and chart.
However, the participant thought that it was an error of the system for not detecting the time range in the chart and moved on.
P6, on the other hand, made a typo by typing in the year 1987 instead of 1997, causing \toolname{} to detect a time range different from what the participant intended (Figure~\ref{fig:results}b Sentence 4)
As with the other participant, the participant deemed it an error of our tool and continued on.
We believe that these errors are partially due to the authors trusting themselves more than \toolname{} and that such issues will diminish as \toolname{} becomes more accurate.
\section{Limitations and Future Work}

\noindent\textbf{\textit{Providing additional guidance in authoring charts and captions.}}

Our tool currently provides rudimentary guidance in authoring charts and captions with matched emphasis but can be improved to provide further guidance to support authors.
As P3 and P8 suggested, the tool can be extended to provide more information about the detected features (e.g., if the feature is rising/falling, the slope).
Future work could also apply natural language generation techniques and suggest automatically generated captions for selected prominent features to further help with the authoring process.
Alternatively, future work could provide further explanations and revision suggestions on detected emphasis mismatches or factual errors to help authors make more informed decisions.
Furthermore, during the pre-study activity of choosing the most effective caption and citing rationale, four of the participants mentioned that useful integration of external information not available in the chart is a property of effective captions, which is in line with the findings of Kim et al.~\cite{kim2021towards}.
Based on these observations, we believe that suggestion of suitable external information in the process could also provide useful additional guidance.
Another potential future direction would be to extend the tool to assist in writing chart descriptions for accessibility. As blind and sighted individuals perceive the usefulness of chart descriptions differently~\cite{lundgard2021accessible}, properties of effective text descriptions for blind individuals should be incorporated into the extended tool design.

\vspace{0.05in}
\noindent\textbf{\textit{Avoiding introducing bias to chart-caption pairs.}}
While all participants agreed on the usefulness of the tool, some participants suggested that guidance needs to be provided to the authors with careful consideration.
P8's concern about the potential biasing of the chart and caption was shared by two other participants.
They mentioned that being shown the prominent features too early while exploring the data could skew what messages they decide to include in the caption.
From these comments, we believe that the guidance provided by the tool could be personalized and adjusted based on the state of the writing process, avoiding too much guidance during the exploratory stages.
Moreover, focusing only on the match between chart and caption emphasis can lead to biased messages or violations of principles of good visualization design, such as the banking to 45 degrees principle~\cite{cleveland1988shape}.
Future developments of the tool can include detectors for biased representation of data and violations of good visualization design to help authors stay faithful to the data within the boundaries of good visualization design while matching chart and caption emphasis.

\vspace{0.05in}
\noindent\textbf{\textit{Generalizing tool to other chart features \& chart types.}}
Our tool is designed to work with time-series line charts, with a focus on universally present features: local extrema and trends. 
Yet, depending on the domain and context of the data, time-series charts can include other types of features, such as seasonal or cyclic patterns in monthly sales of AC units, L- or V-shapes in curves showing economic recovery, or double bottoms in stock charts (`W'-shaped features signaling potential future increases in stock price).
We believe that future work can expand \toolname{} by implementing such features for authors in specific domains and contexts.
Furthermore, future work can go beyond univariate time-series line charts and cover other chart types, starting with the more common charts~\cite{battle2018beagle} such as multi-line charts, bar charts, and scatter plots.
However, we expect that the generalization of \toolname{} to other chart features and chart types will require a deeper understanding of the features not currently covered in this work. \toolname{} is based on being able to compare the prominence of the chart features, and generalization of the tool will require similar levels of understanding of the chart features not covered in our work.
Moreover, the design of the tool is grounded on the assumption that there is a one-to-one mapping between the x-axis and the data points. While the current design would hold for chart types that follow this assumption (e.g., horizontal bar charts), the design of the tool may need modifications for other chart types to help users recognize features while avoiding visual clutter.

\vspace{0.05in}
\noindent\textbf{\textit{Incorporating information outside sentence-chart pairs.}}
Our text references extractor operates at the sentence level and utilizes information available strictly within the sentence and the chart.
The text references extractor currently does not utilize the information available in prior sentences. 
For instance, our tool fails to detect any references in Figure~\ref{fig:results} (d) Sentence 3 because the reference to time (\textit{`May 2012'}) appears in the previous sentence.
Anaphora resolution methods~\cite{mitkov2014anaphora} would be a starting point for future work. 
In addition, while our use of the Named Entity Module in the Stanford CoreNLP toolkit~\cite{chang2012sutime,finkel2005incorporating,manning2014stanford} allows our text references extractor to cover a variety of time expressions (e.g., year, date, time, seasons, quarters, duration, etc.), it is unable to comprehend time expressions requiring external knowledge.
For example, were it not for the explicit mention of the year \textit{`2020'} in Figure~\ref{fig:results} (c) Sentence 3, the tool would not have been able to comprehend the time range described as \textit{`the period of Covid-19.'}
Allowing the text references extractor to draw information from external knowledge bases (e.g., WolframAlpha~\cite{wolfram2022wolframalpha}) or LLM-based tools (e.g., GPT~\cite{brown2020language}) could allow it to detect more descriptions of time reliably.

\vspace{0.05in}
\noindent\textbf{\textit{Computing visual prominence in rendered charts.}}
Our tool, like that of Hullman et al.~\cite{hullman2013contextifier}, utilizes only the properties of the underlying time-series data to approximate visual prominence; the tool does not consider the rendering of the chart.
But, chart authors often use specific encodings (e.g., color, size) and annotations that guide the readers' attention to specific chart features. Such encodings and annotations can affect the visual prominence of features. Future work could analyze the visual properties of the rendered chart as well as the statistical properties of the underlying data to develop 
additional measures of visual prominence.

\vspace{0.05in}
\section{Conclusion}
Writing caption text whose emphasis matches chart emphasis is important for communicating the intended message to the reader.
Through a survey of chart-caption pairs in the real world, we find that there is often a mismatch between chart emphasis and caption emphasis. To address this issue, we introduce \toolname{}, a tool to help authors align the message in the caption text with the prominent features in the chart.
The tool comprises the \textit{time-series prominent feature detector} that identifies chart emphasis and performs at the level of prior crowdsourcing algorithm and the \textit{text reference extractor} that extracts text emphasis and matches them with chart emphasis. Based on feedback we collected from users, \toolname{} is both useful and easy to use for writing effective captions. We identify future directions in this space for further supporting chart authors with better caption-writing tools.
Our code is available at: \url{https://github.com/dhkim16/EmphasisChecker-release}.









\acknowledgments{
This work is supported by the Brown Institute for Media Innovation.
}
	

\bibliographystyle{abbrv-doi-hyperref}

\bibliography{bibliography}

\end{document}